\documentclass[journal=acsami,manuscript=article,layout=twocolumn]{achemso}
\usepackage[utf8]{inputenc}

\DeclareUnicodeCharacter{2082}{$_2$}
\usepackage{chemformula} 
  
\usepackage[version=3]{mhchem} 
\usepackage{hyperref}
\usepackage{xr-hyper} 
\usepackage{graphicx}
\usepackage{multirow}
\usepackage{soul}
\usepackage{subcaption} 
\usepackage{cuted}
\usepackage[none]{hyphenat}
\usepackage{subfiles}
\usepackage{afterpage}

\usepackage{caption}     

\usepackage[T1]{fontenc}
\usepackage{tgheros} 
\DeclareCaptionFont{arialish}{\fontfamily{qhv}\selectfont}

\usepackage{titlesec}
\titleformat{\section}{\normalfont\bfseries\Large}{\thesection.}{0.5em}{}
\titleformat{\subsection}{\normalfont\bfseries\large}{\thesubsection}{0.5em}{}
\titleformat{\subsubsection}{\normalfont\bfseries}{\thesubsubsection}{0.5em}{}

\DeclareCaptionLabelFormat{period}{#1~#2. }
\usepackage[version=3]{mhchem} 

\author{Jeong Min Choi}
\affiliation[SNU]
{Department of Materials Science and Engineering and Research Institute of Advanced Materials, Seoul National University, Seoul 08826, Republic of Korea}
\altaffiliation{These authors contributed equally to this work}
\author{Jaehoon Kim}
\affiliation[SNU]
{Department of Materials Science and Engineering and Research Institute of Advanced Materials, Seoul National University, Seoul 08826, Republic of Korea}
\altaffiliation{These authors contributed equally to this work}
\author{Ji-Hwan Lee}
\affiliation[SNU]
{CSE Team, Semiconductor R\&D Center, Samsung Electronics, Hwaseong 18448, Republic of Korea}
\author{Won-Joon Son}
\affiliation[SNU]
{CSE Team, Semiconductor R\&D Center, Samsung Electronics, Hwaseong 18448, Republic of Korea}
\author{Seungwu Han}
\affiliation[SNU]
{Department of Materials Science and Engineering and Research Institute of Advanced Materials, Seoul National University, Seoul 08826, Republic of Korea}
\alsoaffiliation[KIAS]
{Korea Institute for Advanced Study, Seoul 02455, Republic of Korea}
\email{hansw@snu.ac.kr}
\phone{+82 (02)880 1541}

\title[An \textsf{achemso} demo]
  {Atomistic Insights into Cu/amorphous-Ta$_x$N Interfacial Adhesion via Machine Learning Interatomic Potentials: Effects of Stoichiometry and Interface Construction}

\abbreviations{Ta$_x$N,a-Ta$_x$N,DFT,MD,MLIP,SMD,VASP,PBE,AIMD,ICSD}

\let\oldmaketitle\maketitle
\let\maketitle\relax
\begin{document}







\twocolumn[
 \begin{@twocolumnfalse}
  \oldmaketitle
   \begin{abstract}

Accurate understanding and control of interfacial adhesion between Cu and Ta$_x$N diffusion barriers are essential for ensuring the mechanical reliability and integrity of Cu interconnect systems in semiconductor devices. Amorphous tantalum nitride (a-Ta$_x$N) barriers are particularly attractive due to their superior barrier performance, attributed to the absence of grain boundaries. However, a systematic atomistic investigation of how varying Ta stoichiometries influences adhesion strength at Cu/a-Ta$_x$N interfaces remains lacking, hindering a comprehensive understanding of interface optimization strategies. In this study, we employ machine learning interatomic potentials (MLIPs) to perform steered molecular dynamics (SMD) simulations of Cu/a-Ta$_x$N interfaces. We simultaneously evaluate three distinct interface construction approaches—static relaxation, high-temperature annealing, and simulated Cu deposition—to comprehensively investigate their influence on adhesion strength across varying Ta compositions ($x=1, 2, 4$). Peak force and work of adhesion values from SMD simulations quantitatively characterize interface strength, while atomic stress and strain analyses elucidate detailed deformation behavior, highlighting the critical role of interfacial morphologies. Additionally, we explore the atomistic mechanisms underlying cohesive failure, revealing how targeted incorporation of Ta atoms into Cu layers enhances the cohesive strength of the interface. This study demonstrates how MLIP-driven simulations can elucidate atomic-scale relationships between interface morphology and adhesion behavior, providing insights that can guide future atomistic engineering strategies toward enhancing intrinsic barrier adhesion, potentially enabling liner-free interconnect technologies.

   \end{abstract}
  \end{@twocolumnfalse}
\vspace{\baselineskip}]

\clearpage 


\section{Introduction}

In semiconductor interconnect systems, Cu is widely used for metal lines due to its superior electrical conductivity and higher resistance to electromigration compared to aluminum and its alloys \cite{8268387}. However, a critical challenge arises when Cu contacts dielectric materials directly, as Cu diffusion can form defect phases like Cu$_3$Si, degrading electrical properties \cite{10.1116/1.575620,10.1063/1.345194}. To mitigate this issue, Ta$_x$N is commonly used as a diffusion barrier due to its excellent barrier performance and thermodynamic stability against Cu diffusion \cite{10.1116/1.588818,10.1116/1.1495906,10.1143/jjap.47.6953,10.1146/annurev.matsci.30.1.363,10.1063/1.350566}. 

Another major issue in Cu interconnect technology is the poor adhesion between Cu and adjacent materials, potentially leading to delamination and Cu agglomeration. To address this challenge, liner materials such as tantalum, cobalt, and ruthenium are often introduced between the Cu and the diffusion barrier \cite{10.1109/iitc.2001.930001,10.1149/2.009312jes,10.1109/iitc.2006.1648684}. However, the push toward highly integrated circuits and thinner interconnect structures motivates the exploration of liner-free designs, emphasizing the intrinsic adhesion properties of the barrier materials themselves \cite{10.1149/1.3267881}. Ta$_x$N has been reported to exhibit strong adhesion and enhanced wettability with Cu, as the Ta/N ratio increases \cite{10.1149/1.3267881,10.1116/1.1926289,10.1143/JJAP.47.1042}. In addition, Ta$_x$N is expected to form an amorphous structure at a reduced thickness \cite{10.1116/1.581697,10.1116/1.1495906,10.1063/1.373566}, and such amorphous-Ta$_x$N (a-Ta$_x$N) structures are known to feature superior barrier properties due to the absence of grain boundaries \cite{10.1143/jjap.47.6953,10.1063/1.373566,10.1116/1.582166}. Despite these promising characteristics, the direct experimental measurement of adhesion and structural properties remains challenging because of the thinness of the interconnect systems \cite{10.1146/annurev-matsci-080819-123640}. 

Motivated by the technological significance of Cu/Ta$_x$N interfaces, computational approaches like density functional theory (DFT) and classical molecular dynamics (MD) have been extensively employed.
DFT studies have analyzed adhesion energies, diffusion, and electronic properties, providing fundamental insights into interfacial behavior \cite{10.1103/physrevb.79.214104,10.1039/c8cp01839a,10.1103/physrevb.76.245434,10.1016/j.tsf.2004.06.176,10.1063/1.3369443}. Meanwhile, classical MD simulations have explored the structural evolution and diffusion characteristics of these interfaces under different conditions, including crystalline Cu \cite{10.1063/1.4997677,10.1063/1.1630353}, liquid Cu \cite{10.1016/j.commatsci.2014.01.028}, and amorphous Cu-Ta mixtures \cite{10.1063/1.4905103}. Furthermore, Cu growth simulations on Ta substrates have facilitated understanding of the microstructural development in thin Cu films \cite{PhysRevB.81.045410,10.1557/proc-721-j2.3,10.1023/b:jcad.0000036802.46424.ee,10.1143/jjap.51.06ff14}. MD studies employing reactive potentials have particularly emphasized the atomic-scale sensitivity of interface adhesion. For instance, Ref.~\citenum{10.1021/acsami.4c03418} demonstrated that the interfacial oxygen content significantly influences Cu/SiO$_2$ adhesion through stabilization of Cu--O bonds and modifications in the local bonding environment. 

Nevertheless, each approach has limitations. DFT calculations are limited by the size of the system, restricting atomistic analyses of failure processes. Classical MD can handle larger scales, but its accuracy depends critically on the empirical potentials used, which are often unavailable or inaccurate for mixed bonding environments like Cu/Ta$_x$N interfaces.

Recently, machine learning interatomic potentials (MLIPs) have emerged as a promising method to bridge the accuracy of DFT and the computational scale of classical MD \cite{10.1002/adma.201902765}. When trained on suitable DFT training set, MLIPs offer near-DFT accuracy at a substantially reduced computational cost, enabling reliable simulations of large, complex interfacial systems \cite{10.1080/27660400.2023.2269948,10.1063/5.0244175}. 

Indeed, recent MLIP studies have consistently demonstrated how detailed atomic-scale interface configurations significantly influence adhesion properties. Ref.~\citenum{10.1038/s41598-023-44265-6} employed moment tensor potential\cite{doi:10.1137/15M1054183} based simulations to investigate various interfaces, highlighting that surface termination significantly impacts adhesion behavior; W/TiN interfaces with Ti termination exhibited purely adhesive failures, while N-terminated surfaces led to cohesive failure within the W layer, substantially increasing pull-off strength. Similarly, their studies on Ru/SiO$_2$ interfaces \cite{10.1039/d4tc04870a} revealed significant variations in the adhesion energy linked to interface stoichiometry, emphasizing the critical role of atomistic features such as Ru--O--Si bridge bonds. Ref.~\citenum{10.1021/acsami.4c06055}, using neuroevolution-potential \cite{PhysRevB.104.104309}, also underscored atomic-level factors like annealing and strain engineering at AlN/diamond interfaces, demonstrating marked improvements in adhesion through enhanced atomic bonding stability. Additionally, Ref.~\citenum{10.1016/j.apsusc.2025.162558} utilized CHGNet\cite{10.1038/s42256-023-00716-3} to explore Cu/TaN interfaces under different biaxial strains, identifying significant strain-induced differences in adhesion strength and ductility influenced by crystallographic orientation and surface chemistry. 

Although previous computational studies underline the critical impact of atomic-scale interface configurations on adhesion properties, the majority of them still predominantly employ simplified interface construction methods, typically involving short annealing at room temperature \cite{10.1038/s41598-023-44265-6, 10.1039/d4tc04870a, 10.1016/j.apsusc.2025.162558, 10.1021/acsami.4c03418}. Such studies usually consider interface stoichiometry or construction methods independently, limiting insights into how these factors collectively influence adhesion strength and fracture behavior. Thus, a comprehensive approach that combines variations in the stoichiometry of the interface and various interface construction methods could provide deeper understanding.

Experimental studies further highlight the critical influence of detailed interface construction on mechanical performance. For example, Ref.~\citenum{10.1116/1.3602079} demonstrated that optimized physical vapor deposition conditions, enhancing Cu flux ionization, significantly improved Cu/Ta-based interface uniformity and adhesion by reducing interfacial roughness and promoting atomic-level chemical homogeneity. Similarly, Ref.~\citenum{JEONG2025112307} reported that variations in sputtering techniques significantly influenced the interfacial microstructure and adhesion strength of Cu--Mn/Ta/TaN interfaces.

Motivated by the aforementioned discussions, we herein employ MLIPs to model the Cu/a-Ta$_x$N interface and investigate its adhesion and fracture behavior under various interfacial configurations using steered molecular dynamics (SMD) simulations, which enable direct modeling of the interface delamination process at finite temperature. In particular, we explore interface construction through static relaxation, high-temperature annealing, and simulated deposition, enabling direct comparisons among different modeling procedures. Furthermore, we evaluate how variations in Ta$_x$N composition ($x=1, 2, 4$) influence the stability of the interface. By leveraging the accuracy and computational efficiency of MLIPs, this study aims to provide deeper insights into the adhesion properties of Cu/a-Ta$_x$N interfaces, contributing meaningfully to the advancement of liner-free interconnect technologies.

\section{Methods}
\subsection{Construction of training set}

\begin{figure*}[ht]
    \centering
    \includegraphics{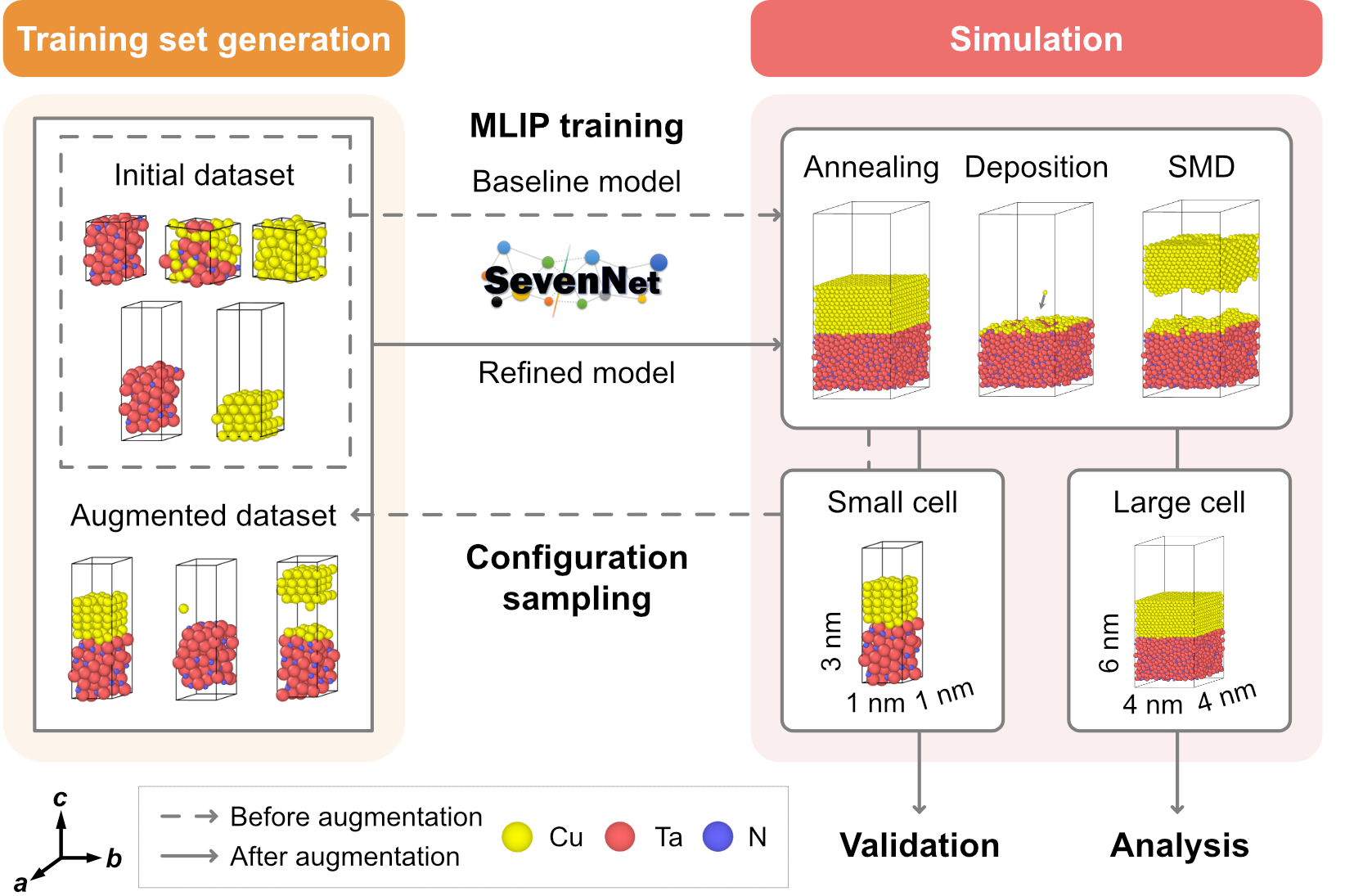}
    \caption{Schematic workflow of the study. The initial training set is generated by DFT and used to train a baseline MLIP with the SevenNet package. The baseline MLIP is refined by incorporating augmented configurations obtained from MLIP MD simulations of annealing, deposition, and SMD processes at Cu/a-Ta$_x$N interfaces in small simulation cells ($3\mathord{\times}1\mathord{\times}1$ nm$^3$). The refined MLIP is validated on small cells and subsequently applied to large-scale simulations ($6\mathord{\times}4\mathord{\times}4$ nm$^3$) for interface analysis. Cu atoms are shown in yellow, Ta in red, and N in blue. Dashed arrows denote processes before augmentation, and solid arrows denote process after augmentation.}
    \label{fig:schematic_workflow}
\end{figure*}

\begin{table*}[htbp]
\centering
\caption{Overview of the training set configurations.}
\label{tab:training_set}
{
\begin{tabular}{llll}
\hline
\textbf{Dataset} & \textbf{Type} & \textbf{Description} & \textbf{Structures}\\
\hline
Initial & bulk & strained crystal polymorphs & 441\\
       & bulk & FCC Cu annealing (500 K) & 125\\ 
       & bulk & liquid and a-Ta$_x$N (MQA) & 2,313\\
       & slab & FCC Cu$(111)$ annealing (600 K) & 122\\
       & slab & a-Ta$_x$N annealing (1000 K) & 250\\
       & bulk & liquid and a-Cu (MQA) & 770\\
       & bulk & liquid and a-Ta$_x$NCu$_y$ (MQA) & 3,104\\
Augmented & interface & Cu/a-Ta$_x$N annealing (1000 K) & 198\\
         & interface & Cu/a-Ta$_x$N SMD (300 K) & 294\\
         & interface & Cu deposition on a-Ta$_x$N slab (700 K) & 367\\
\hline
\end{tabular}
}
\end{table*}

\begin{figure*}[ht]
    \centering
    \includegraphics{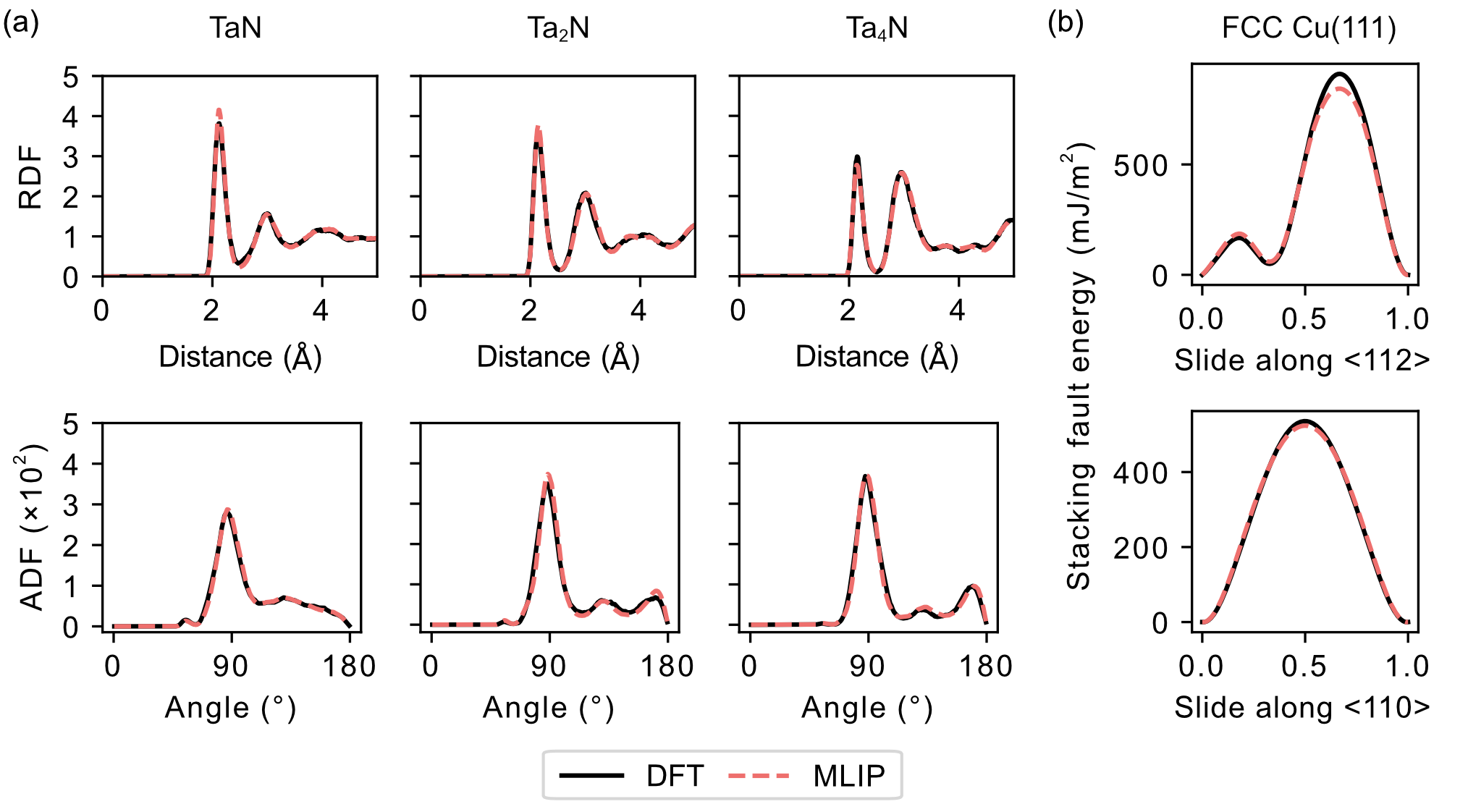}
    \caption{Validation of the refined MLIP model (red dashed lines) against DFT reference data (black solid lines). (a) RDF (top) and ADF (bottom) of a-Ta$_x$N, with compositions TaN, Ta$_2$N, and Ta$_4$N shown from left to right. (b) GSF energy curves for FCC Cu$(111)$ under sliding along the $\langle112\rangle$ (top) and $\langle110\rangle$ (bottom) directions.}
    \label{fig:val_rdfadf_SFE}
\end{figure*}

All DFT calculations are performed using the Vienna Ab initio Simulation Package (VASP) \cite{10.1016/0927-0256(96)00008-0,10.1103/physrevb.54.11169,10.1103/physrevb.59.1758,10.1016/0022-3093(95)00355-x}. The exchange-correlation functional is described using the Perdew-Burke-Ernzerhof (PBE) \cite{10.1103/physrevlett.77.3865} form of the generalized gradient approximation. To ensure convergence of total energy and atomic forces to within 10 meV/atom and 0.2 eV/\AA{}, respectively, we employ the following settings for all single-point calculations: a plane-wave cutoff energy of 450 eV and a $2\mathord{\times}2\mathord{\times}2$ \textbf{k}-point grid for lattice parameters approximately 10 \AA{} in size, and a self-consistent convergence criterion of $10^{-6}$ eV.

To build training set for the MLIPs, ab initio molecular dynamics (AIMD) simulations are first carried out using less stringent computational settings: a cutoff energy of 400 eV, $\Gamma$-point sampling for the \textbf{k}-point grid, and a self-consistent convergence criterion of $10^{-4}$ eV. A timestep of 2 fs is used, and snapshots are sampled at specific time intervals. Each sampled structure is then subjected to single-point calculation with more stringent DFT settings described above, and subsequently included in the training set.

As illustrated in \autoref{fig:schematic_workflow}, the training set consists of two parts: (1) an initial dataset and (2) an augmented dataset. The initial dataset includes crystalline polymorphs, amorphous bulk, and surface structures. For all amorphous bulk and surface a-Ta$_x$N structures, Ta/N ratios of 2 and 4 ($x = 2, 4$) are included.

Polymorphs of Ta$_x$N and Cu, including their stable and metastable phases, are obtained from the Inorganic Crystal Structure Database (ICSD) \cite{10.1080/08893110410001664882} and relaxed using DFT. To explore different structural states, each structure is volumetrically strained from $-$5\% to $+$5\% in increments of 0.5\%.

AIMD simulation for face-centered cubic (FCC) Cu is run in the NVT ensemble at 500 K using a 64-atom supercell. The simulation lasts 5 ps, with structures sampled every 40 fs. 

The amorphous phases of Ta$_x$N are generated via melt-quench-anneal (MQA) AIMD. Approximately 100 atoms are initially randomized in a cubic cell and superheated at 6000 K for 6 ps. The systems are melted at 5000 K for 10 ps, quenched to 300 K at a rate of $-$100 K/ps, and annealed at 500 K for 10 ps. The melting and annealing stages use the NVT ensemble, while quenching is performed in the NPT ensemble. Snapshots are taken every 40 fs for melting and annealing and every 100 fs for quenching. The a-Ta$_x$N slabs are prepared by cleaving the bulk structures and fixing the bottom 5 \AA{}. These are equilibrated for 1 ps and then simulated at 1000 K for 10 ps in the NVT ensemble, sampling every 40 fs. 

The $(111)$ plane of FCC Cu, known as the preferred surface for interconnect systems with Ta$_x$N \cite{10.1149/1.1531974}, is selected as the initial structure of the Cu surface. The slab consists of 80 atoms in 5 layers, with the bottom two layers fixed. AIMD is performed at 600 K for 5 ps in the NVT ensemble, with configurations sampled every 40 fs.

To represent the disordered Cu structures that may arise during interface annealing, deposition, and SMD simulations, amorphous Cu is also generated using similar MQA procedure, except melting is performed at 2000 K and quenching employs the NVT ensemble instead of NPT. 
Additionally, to capture complex interactions at the Cu/a-Ta$_x$N interface, ternary amorphous systems are generated with compositions, Ta$_{36}$N$_{18}$Cu$_{64}$ ($x = 2$), Ta$_{48}$N$_{12}$Cu$_{64}$ ($x = 4$), and Ta$_{72}$N$_{18}$Cu$_{12}$ ($x = 4$). These follow the same MQA procedure for amorphous Cu, except melting at 5000 K, with 100 fs sampling for $x = 4$ to reduce computational cost.

The augmented dataset is generated using structures from simulations driven by a baseline MLIP trained on the above initial dataset. (We employ the training method as detailed in the next section.) This includes configurations sampled from annealing, SMD, and deposition at small cell from Cu/a-Ta$_x$N interfaces (see \autoref{fig:schematic_workflow}). All sampled configurations are recomputed with DFT using the same high-accuracy settings employed for the initial dataset.

To generate the lower part of the interface, a-Ta$_x$N structures are constructed on the lattice of $4\mathord{\times}4$ FCC Cu$(111)$ surfaces relaxed with the basline MLIP. The same MQA protocols are used as those applied during a-Ta$_x$N bulk generation, except for performing quenching in the NP$_z$T ensemble to allow only vertical relaxation, and then vacuum layers are introduced to construct slab structures.

The resulting a-Ta$_x$N slabs are used in two types of interface simulations. In the first approach, Cu and a-Ta$_x$N slabs are adjoined to form an interface, which is equilibrated at 1000 K for 1 ns, followed by quenching to 300 K at a rate of $-$10 K/ps and annealing at 300 K for 100 ps, while the bottom half of the a-Ta$_x$N slab is kept fixed. Configurations are sampled every 10 ps during the equilibrating stage at 1000 K. SMD simulations are subsequently applied to these annealed interfaces, and approximately 300 configurations are extracted, covering the entire trajectory up to the point where the interface is fully separated. In the second approach, Cu atoms are deposited onto the a-Ta$_x$N slabs. During deposition, the bottom half of the a-Ta$_x$N slab is thermostatted at 700 K in the NVT ensemble, while the upper part of the slab evolves without thermostat. A total of $\sim$60 Cu atoms are deposited and snapshots are collected every 400 fs. Detailed deposition conditions are provided in Section 2.3.

All MLIP-based MD simulations in this study are performed using LAMMPS software \cite{THOMPSON2022108171}. For bulk phase simulations, both the NVT and NPT ensembles employ the Nosé-Hoover thermostat and barostat\cite{1984MolPh..52..255N}. All interface simulations, including annealing, SMD, and deposition in the NVT ensemble, use the Langevin thermostat \cite{PhysRevB.17.1302} for temperature control. The training sets are categorized in \autoref{tab:training_set}, comprising a total of 843,910 training points: 420,169 for Ta, 142,578 for N and 281,163 for Cu. Visualization and analysis of all atomic structures are performed using OVITO software \cite{Stukowski_2010}.

\begin{figure}[t!]
    \centering
    \includegraphics{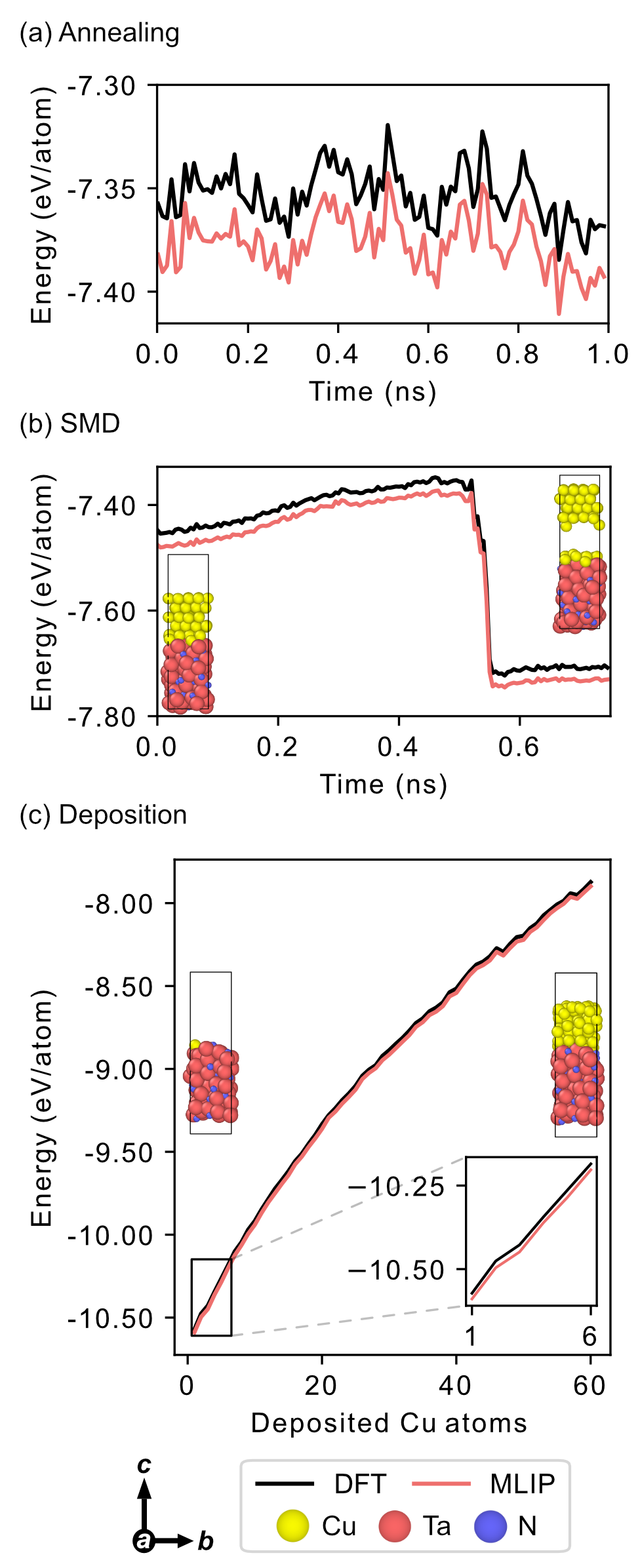}
    \caption{Comparison of DFT (black) and MLIP (red) energies along MLIP MD trajectories for TaN. (a) Annealing of the Cu/a-TaN interface, (b) SMD simulation of Cu/a-TaN, (c) Cu deposition on an a-TaN slab.}
    \label{fig:val_simulation_E}
\end{figure}

\begin{figure*}[ht]
    \centering
    \includegraphics{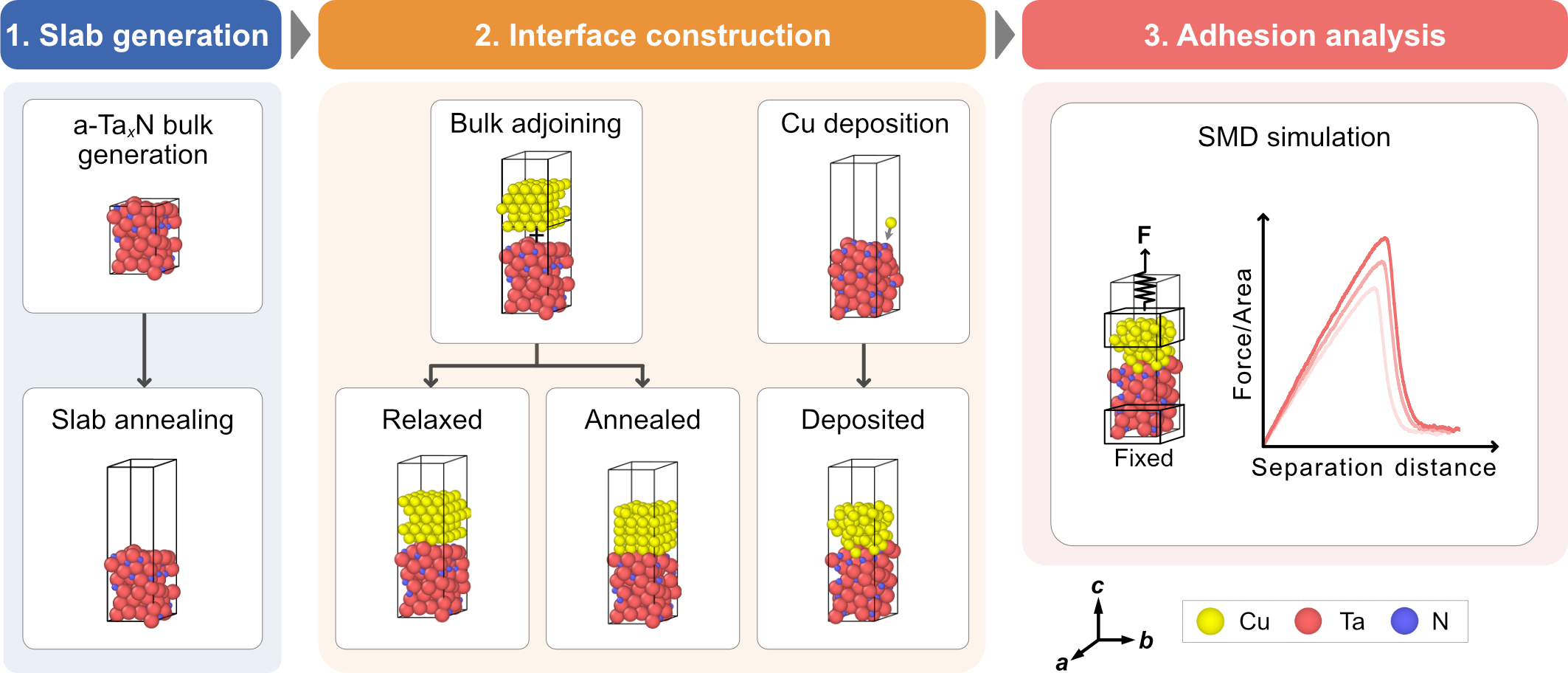}
    \caption{Schematic workflow for interface construction and adhesion analysis. (1) a-Ta$_x$N slabs are generated using the MQA procedure, followed by vacuum introduction and annealing at 1000 K. (2) Three interface construction methods are considered: bulk adjoining with either static relaxation or annealing at 1000 K, and simulated deposition of Cu atoms onto a-Ta$_x$N slabs. (3) The resulting interfaces are subjected to SMD simulations, where the Cu slab is pulled away from the fixed a-Ta$_x$N slab to evaluate interfacial adhesion. Cu atoms are shown in yellow, Ta in red, and N in blue.}
    \label{fig:schematic_interface_modeling}
\end{figure*}

\subsection{Training and validation of MLIP models}

The DFT calculations for the training data result in 7,984 structures in total. The training data are randomly shuffled and split into training and validation set in a 9:1 ratio. We base our MLIP implementation on SevenNet package\cite{10.1021/acs.jctc.4c00190}, utilizing its available NequIP architecture, an E(3)-equivariant message-passing graph neural network \cite{10.1038/s41467-022-29939-5}. The MLIP used in this study features channel counts of 32, and message-passing layer counts of 3 along with a maximum degree of representation ($l_{max}$) of 3 and a cutoff radius of 5.5 \AA{}. For faster and more memory-efficient training and inference, the FlashTP\cite{leeflashtp} CUDA kernel is utilized in the convolution layers of SevenNet, providing roughly a fourfold increase in throughput. After training MLIP, incorporating augmented configurations, the refined MLIP demonstrates the values of root mean square error (RMSE) of 4.5 meV/atom, 0.3 eV/\AA{}, and 7.1 kbar for the energy, forces, and stress components, respectively, in the validation set.

Once the MLIP has been trained, validations of the MLIP are performed using an external test set that is not included in the training or validation set. First, the ability of MLIP to accurately reproduce the structural properties of amorphous phases of Ta$_x$N is rigorously assessed. In \autoref{fig:val_rdfadf_SFE}a, the radial distribution functions (RDF) and angular distribution functions (ADF) of a-Ta$_x$N generated by the MLIP are compared with those obtained from DFT calculations. Ten amorphous structures for each composition are sampled by MLIP, calculated with identical DFT protocols, and then annealed at 300 K to obtain averaged RDF and ADF values. These results agree well with DFT reference data. Notably, the positions and heights of the a-TaN peaks are well reproduced, even though these configurations are not explicitly included in the training set. 

To robustly account for possible stacking faults introduced during interface construction and subsequent delamination processes, generalized stacking fault (GSF) energy surfaces are computed along the $(111)$ plane in FCC Cu using both DFT and MLIP, as shown in \autoref{fig:val_rdfadf_SFE}b. Despite the absence of explicit Cu stacking fault configurations in the training set, the MLIP accurately reproduces the GSF energy profiles corresponding to displacements along the $\langle112\rangle$ and $\langle110\rangle$ directions on the $(111)$ plane.

Finally, we perform MLIP-MD simulations using the same methodology described in Section 2.1 for the augmented dataset, followed by DFT single-point calculations. This validation process confirms the reliability of MLIP in simulating interface annealing, SMD, and deposition processes. \autoref{fig:val_simulation_E} and Figure S1 present the DFT and MLIP energies along the MD trajectories. Although a constant energy shift in interface annealing MD is observed--23 meV/atom for TaN, 8 meV/atom for Ta$_2$N, and 9 meV/atom for Ta$_4$N--the force components show strong agreement, with mean absolute errors (MAEs) below 0.3 eV/\AA{} (presented in Figure S2), ensuring stable and accurate MD simulations.

\subsection{Interface construction}

In this study, three different methods are employed to prepare the interfaces, and a schematic of the interface construction is illustrated in \autoref{fig:schematic_interface_modeling}. Initially, a-Ta$_x$N structures containing 3,840, 3,072, and 3,200 atoms for $x=1$, $2$, and $4$, respectively, are generated by the MQA procedure. During the quenching step, we employ the NP$_z$T ensemble, consistent with the protocol used to generate the augmented dataset. The lateral lattice parameters, 40.92 \AA{}, are constrained to match the equilibrium dimensions of $16\mathord{\times}16$ FCC Cu$(111)$ surfaces, thereby minimizing the artificial lateral strain upon the formation of the Cu/a-Ta$_x$N interface. Subsequently, the generated a-Ta$_x$N slabs are annealed at 1000 K for 1 ns after introducing a vacuum region. In the first two approaches, annealed a-Ta$_x$N slabs and Cu slabs (consisting of 12 layers) are joined directly with an interfacial distance of 2 \AA{}. While the bottom 5 \AA{} of the a-Ta$_x$N slabs is kept fixed, the first interfaces are then subjected to static relaxation, whereas the second interfaces undergo a subsequent annealing at 1000 K for 1 ns and then quenching to 300 K at a rate of $-$10 K/ps. The third interfaces are generated differently: rather than adjoining bulk slabs, simulated depositions of Cu atoms onto a-Ta$_x$N slabs are performed. The initial velocities of the Cu atoms are sampled from a Maxwell-Boltzmann distribution corresponding to a temperature of 1500 K. Throughout the deposition, the bottom region of the Ta$_x$N slabs, approximately 5 \AA{} in thickness, is fixed. Temperature control is achieved by thermostatting the remaining lower half of the a-Ta$_x$N slab at 700 K under an NVT ensemble, while the upper region is simulated under an NVE ensemble. For each insertion of a Cu atom, the simulation proceeds until the average temperature of nonthermostatted region of the last 1 ps falls below 750 K, preventing excessive heating of the surface. This adaptive procedure results in an average deposition interval of approximately 2 ps per Cu atom. The deposition process continues until a total of 3100 Cu atoms have been introduced, ensuring consistency in the number of Cu atoms compared to the first two interface construction approaches. The cumulative simulation duration for the entire deposition process is approximately 7.7 ns. We systematically compare three interface construction approaches (simply relaxed, annealed, and deposited henceforth). Regardless of the interface construction method employed, all interfaces undergo an final equilibration step at 300 K, prior to performing subsequent SMD simulations.

\subsection{Steered molecular dynamics}
To evaluate the adhesion between Cu and a-Ta$_x$N, we employ SMD simulation, a nonequilibrium method that relates the nonequilibrium work performed during separation to the equilibrium potential of mean force (PMF) via the Jarzynski equality \cite{PhysRevLett.78.2690}.

In our SMD simulations, a virtual spring connects a moving tether point located above the Cu slab to the center of mass of the upper half of the Cu slab. This tether point moves upward at a constant velocity, causing controlled separation at the Cu/a-Ta$_x$N interface. The spring force is proportionally distributed among all Cu atoms in the upper half of the slab based on their atomic masses. Ta and N atoms within 5 \AA{} from the bottom remain fixed.

\begin{figure*}[t!]
    \centering
    \includegraphics{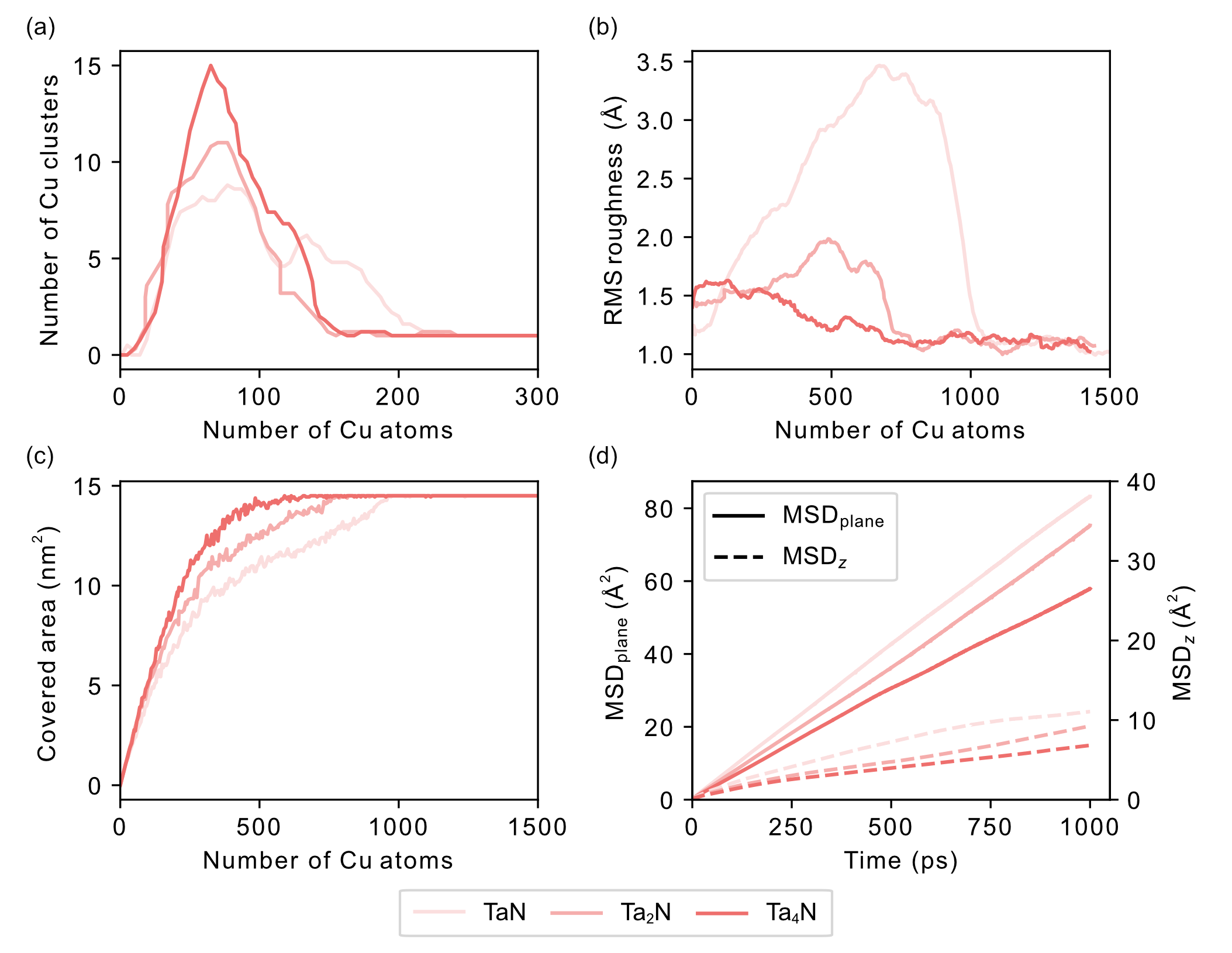}
    \caption{Properties of Cu deposition on a-Ta$_x$N substrates with varying Ta content, with darker lines indicating higher Ta concentration. (a) Number of Cu clusters as a function of deposited atoms, where clusters are defined as groups of two or more atoms with distances below 3.20 \AA{}. (b) Root mean square (RMS) surface roughness during deposition, defined as in Ref.~\citenum{10.1016/j.apsusc.2012.08.082}. (c) Evolution of surface coverage, calculated as the sum of areas of lateral surface bins occupied by at least one Cu atom (bin side length is 2.56 \AA{}, similar to Cu--Cu nearest-neighbor distance). (d) In-plane (solid lines) and out-of-plane (dashed lines) MSD of Cu atoms over time, computed for the first 75 deposited Cu atoms, roughly corresponding to the clustering peaks shown in panel (a).}
    \label{fig:unified_depo}
\end{figure*}

The applied force follows the equation:

\begin{equation}
U_\text{spring} = \frac{1}{2}k \left[ vt - \big( \textbf{R}(t) - \textbf{R}_0 \big) \cdot \textbf{n} \right]^2 \\
\end{equation}
\begin{equation}
\textbf{F}_\text{spring} = -\nabla U_\text{spring}
\end{equation}

where $k$ is the spring constant, $v$ is the pulling velocity, $\textbf{R}(t)$ represents the center of mass position of the pulled Cu atoms at time $t$, $\textbf{R}_0$ is their initial center of mass position and $n$ is the unit vector that defines the pulling direction. The nonequilibrium work $W$ performed in this process is given by:

\begin{equation}
W = \int_{\textbf{r}=\textbf{R}_0}^{\textbf{r}=\textbf{R}_\text{f}} \nabla U_\text{spring} \cdot d\textbf{r}
\end{equation}

where $\textbf{R}_\text{f}$ denotes the final center of mass position of the Cu layers. By averaging the exponential of the work over multiple nonequilibrium simulations, Jarzynski equality allows the determination of the equilibrium PMF\cite{park2004calculating-fbb}, $U_\text{PMF}$:

\begin{equation}
\left\langle \exp(-\beta W) \right\rangle_\text{ensemble} = \exp(-\beta U_\text{PMF})
\label{Jarzynski}
\end{equation}

\begin{figure*}
    \centering
    \includegraphics{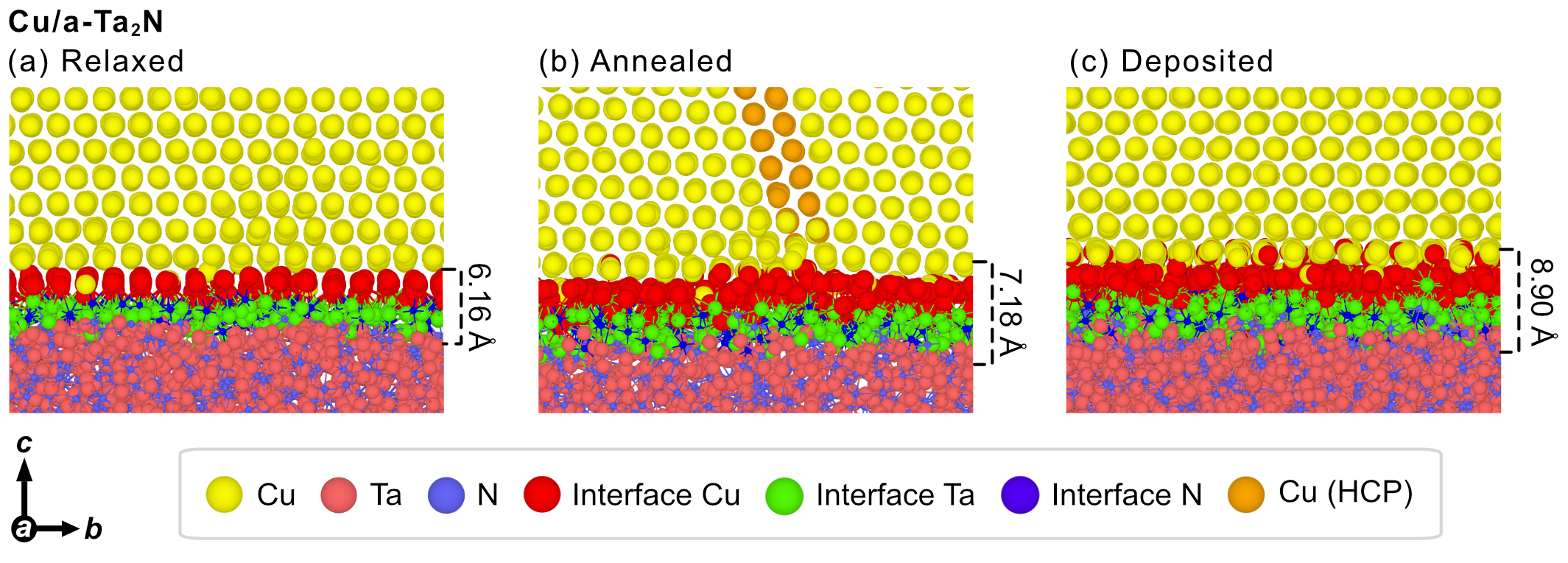}
    \caption{Side-view snapshots of Cu/a-Ta$_2$N interfaces prepared by three different construction methods: (a) relaxed, (b) annealed, and (c) deposited. Dashed lines indicate the thickness of the intermixing region in each interface.
}
    \label{fig:Ta2N_interface}
\end{figure*}

where $\beta = \frac{1}{k_B T}$, $k_B$ is the Boltzmann constant and $T$ is the temperature. In this study, we use the final plateau value of $U_\text{PMF}$ (corresponding to complete interface separation) as the work of adhesion ($W_{\mathrm{ad}}$). In each SMD simulation, we use a spring constant, \(k\), equal to 0.01 eV/\AA{}$^2$/atom, and use a $v$ equal to 0.01 \AA{}/ps. SMD parameter convergence test results are illustrated in Figure S3. For each interface condition, we perform three independent SMD simulations, whose initial configurations are sampled every 100 ps during the final 300 K equilibration MD simulations.

\section{Results} 
\subsection{Deposition process}

\autoref{fig:unified_depo} illustrates how Cu grows on Ta$_x$N by tracking key morphological and dynamical properties. In \autoref{fig:unified_depo}a, the early-stage nucleation peaks at around 75 deposited Cu atoms, after which smaller clusters merge into a dominant island and finally form one Cu cluster. This behavior reflects a Volmer–Weber 3D island growth mode, due to the high nucleation barrier and poor Cu wettability on TaN \cite{Radisic_2003}. The limited number of nucleation sites leads to fewer but larger Cu islands, aligning with studies showing larger Cu grain sizes on TaN compared to Ta under similar conditions \cite{Yang_2003}. As additional Cu atoms are deposited, they preferentially attach to existing clusters, eventually bridging the gaps between islands. Meanwhile, \autoref{fig:unified_depo}b,c show that a higher Ta content in Ta$_x$N fosters a quasi-2D spread of Cu, reducing roughness and accelerating coverage; in contrast, N-rich surfaces are prone to pronounced 3D island formation, leading to patchier, higher-roughness films that risk leaving portions of the barrier surface exposed. If areas of the Ta$_x$N barrier remain uncovered, the later electrochemical plating of Cu can lead to voids or seams because Cu will not deposit uniformly, favoring existing Cu over the bare Ta$_x$N\cite{10.3390/ma13215049}. However, it should be noted that due to the relatively small lateral dimensions of the substrates used in our study ($\sim$4 nm) compared to the device scale, eventually the full coverage of the surface is observed for all Ta$_x$N. Mean squared displacement (MSD) analyses (\autoref{fig:unified_depo}d) reveal that the N-rich substrate exhibits higher mobility in both the in-plane (MSD$_{\text{plane}}$) and out-of-plane (MSD$_{\textit{z}}$) directions, promoting dynamic adatom motion that accelerates 3D island formation. In contrast, Ta-rich films have comparatively lower overall MSD, favoring more confined surface diffusion and flatter, quasi-2D morphologies. Thus, by correlating cluster evolution, roughness, surface coverage, and MSD, \autoref{fig:unified_depo}a-d collectively reveal how interactions between substrate and film govern the transition from 3D island nucleation to quasi-2D film formation. These results align with previous experimental \cite{10.1116/1.1926289,10.1016/j.surfcoat.2004.10.086} and theoretical \cite{SANGIOVANNI2018180,10.1039/d1sc04708f,10.1002/ange.200905360,Aldana_2025} studies and confirm that the systematic trend still holds for the amorphous substrate cases, and further confirm the reliability of the trained MLIP.

\subsection{Interface structures}
\begin{figure}[t!]
    \centering
    \includegraphics[width=\columnwidth]{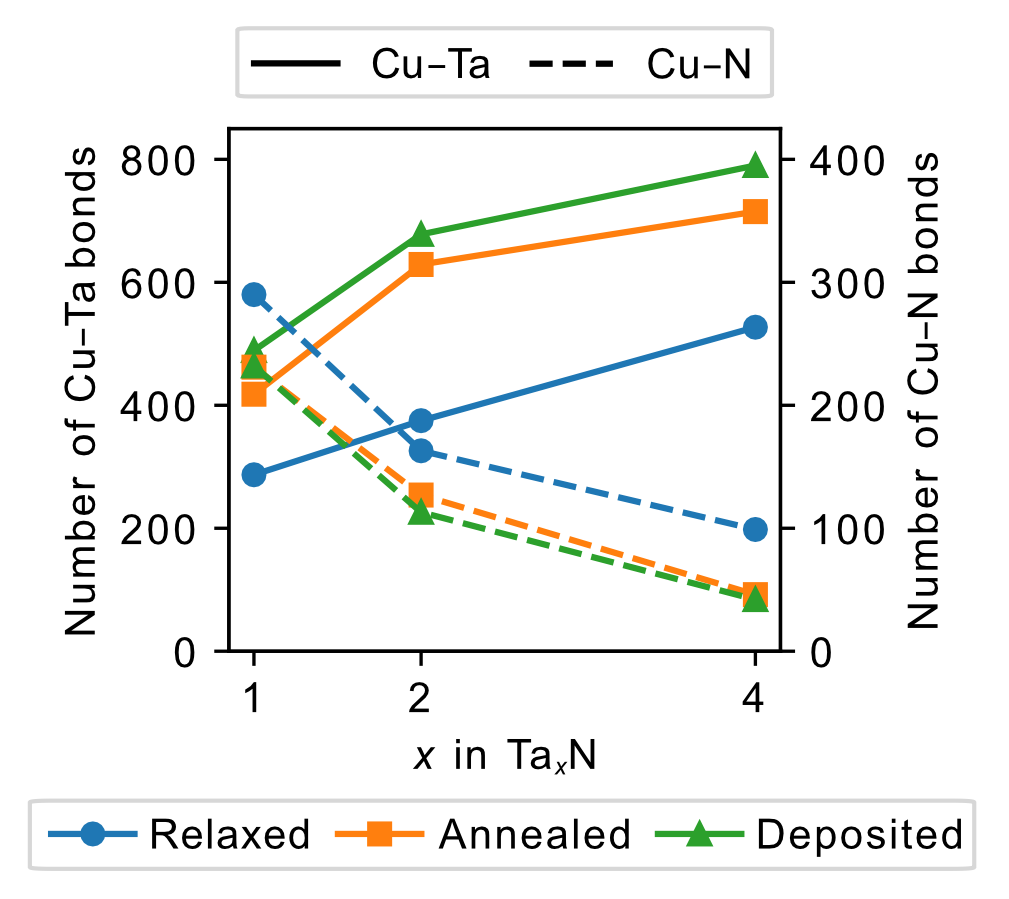}
    \caption{Number of Cu--Ta (solid lines, left axis) and Cu--N (dashed lines, right axis) bonds as a function of Ta$_x$N composition ($x=1,2,4$), categorized by interface construction method: relaxed (blue), annealed (orange), deposited (green).}
    \label{fig:interface_bondnumbers}
\end{figure}

\begin{figure*}[ht]
  \centering
  \includegraphics{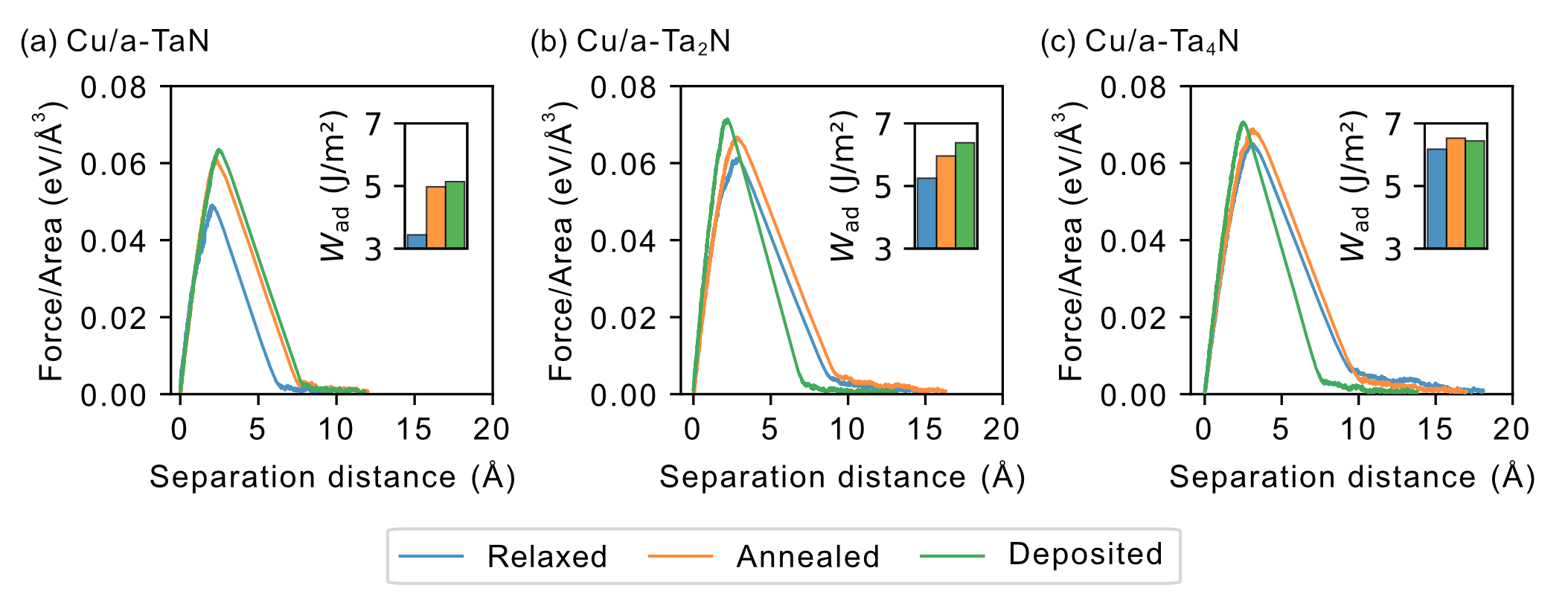}
  \caption{Averaged pulling force as a function of separation distance during SMD simulations for Cu/a-Ta$_x$N interfaces constructed by relaxation (blue), annealing (orange), and deposition (green). (a) Cu/a-TaN, (b) Cu/a-Ta$_2$N, (c) Cu/a-Ta$_4$N. The cross-sectional area of the simulation cell perpendicular to the pulling direction is 14.5 nm$^2$ for all cases. Insets show the corresponding $W_\mathrm{ad}$ obtained from each simulation.}
  \label{fig:smd_pulling_force}
\end{figure*}

\autoref{fig:Ta2N_interface} compares the Cu/a‐Ta$_2$N interfaces generated by three different construction methods (relaxed, annealed, and deposited). We define the intermixing thickness as the vertical distance between the highest and lowest z-coordinates of all atoms that form at least one Cu--Ta or Cu--N bond. Cu--Ta and Cu--N bonds are defined when the interatomic distance is within 3.20 \AA{}. The relaxed interface features the thinnest intermixing region, whereas the annealed interface shows a moderate amount of mixing, and the deposited interface displays the thickest intermixing. This systematic increase in the thickness of the intermixing region from relaxed to annealed to deposited is consistent across all Ta stoichiometries (Figure S4), reflecting the increasingly dynamic nature of atomic rearrangement from static relaxation through thermal annealing to deposition processes.

Although the intermixing thickness provides useful insights, it does not fully capture whether the interfacial bonding network is dense or spatially uniform. To illustrate these differences more clearly, we construct surface mesh images \cite{10.1007/s11837-013-0827-5} (Figure S5) that connect all atoms participating in at least one Cu--Ta or Cu--N bond. These mesh renderings reveal stoichiometry-dependent differences: TaN interfaces show patchy connectivity with significant voids even in annealed and deposited cases, while Ta$_2$N and Ta$_4$N develop extensive, nearly continuous bonding networks under the same conditions. In particular, the mesh density appears similar between Ta$_2$N and Ta$_4$N, suggesting that the bonding network approaches saturation at intermediate Ta contents.

To quantify the interfacial bonding, we compute the number of Cu--Ta and Cu--N bonds in the interfacial region (\autoref{fig:interface_bondnumbers}). Across all construction methods, increasing Ta content systematically enhances Cu–Ta bonding while simultaneously decreasing Cu–N bonding. This opposing behavior demonstrates the preferential bond affinity of Cu for Ta over N \cite{10.1103/physrevb.79.214104, 10.1039/c8cp01839a}. Furthermore, the bond numbers reflect the construction method; the extensively intermixed annealed and deposited interfaces form significantly more bonds than the relaxed interfaces. Finally, the increase in Cu--Ta bonds from Ta$_2$N to Ta$_4$N is less pronounced than from TaN to Ta$_2$N, indicating saturation in Cu--Ta bonding at higher Ta concentrations.

\subsection{Results from SMD simulations}
\subsubsection{Peak forces and work of adhesion}
\begin{figure*}[t!]
    \centering
    \includegraphics{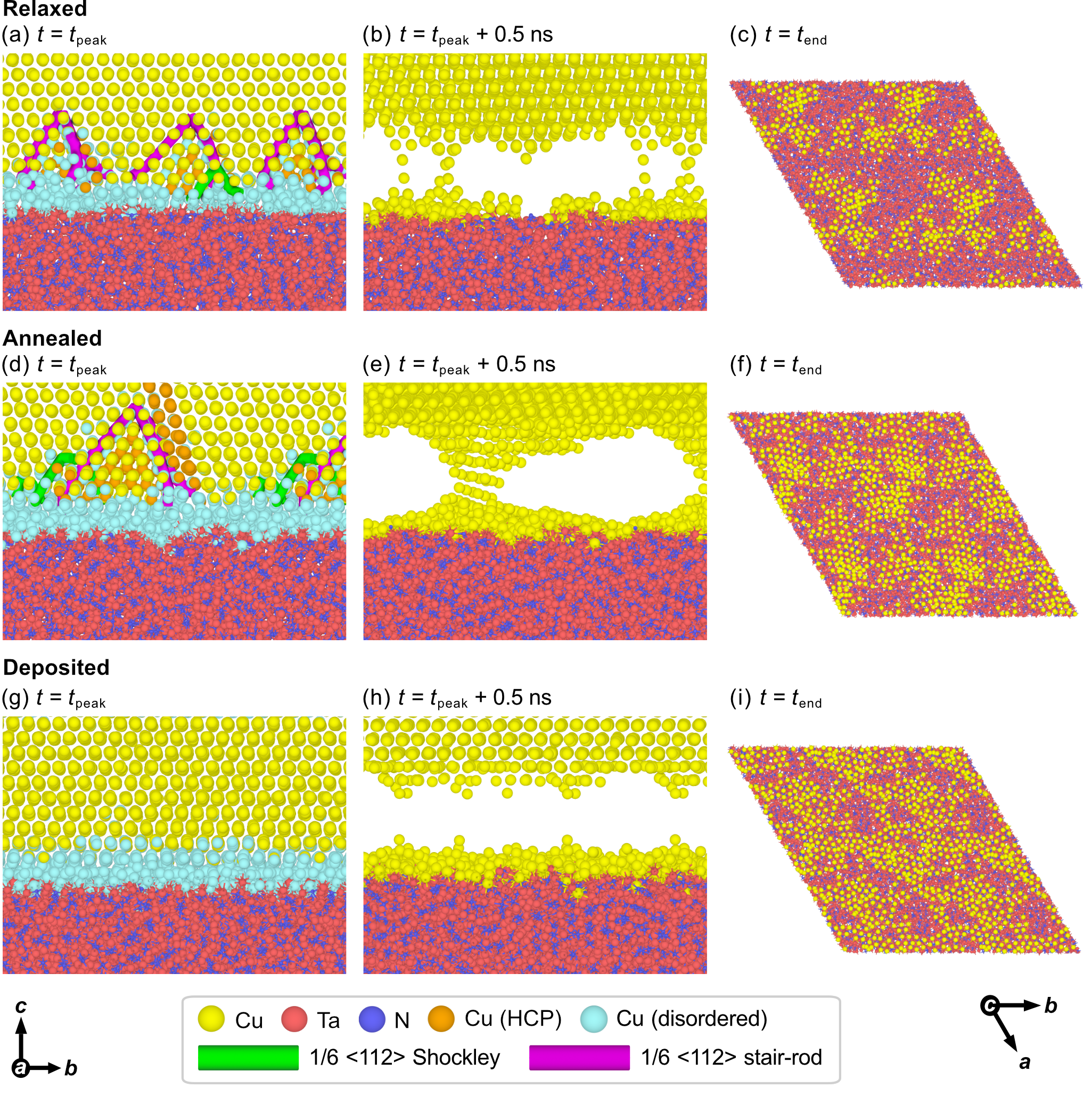}
    \caption{Snapshots from SMD simulations of Cu/a-Ta$_2$N interfaces generated under relaxed (top), annealed (middle), and deposited (bottom) conditions. Columns correspond to different simulation stages: (a,d,g) peak force stage ($t = t_\text{peak}$), (b,e,h) 0.5 ns after peak force stage ($t = t_\text{peak}+0.5\,\text{ns}$), (c,f,i) top view of Ta$_2$N surface when the simulation is ended ($t = t_\text{end}$). For clarity, only the first-column snapshots include structural analysis overlays: dislocation segments identified by the dislocation extraction algorithm (DXA) \cite{Stukowski_2012} are shown as thick tubes (bright-green, 1/6⟨112⟩ Shockley partial dislocations; magenta, 1/6⟨110⟩ stair-rod partial dislocations), and Cu atoms are colored according to their local structural environment (e.g., hexagonal close-packed (HCP) or disordered).}
    \label{fig:SMD_config_Ta2N}
\end{figure*}

To evaluate the adhesion of Cu/a-Ta$_x$N, we perform SMD simulations at 300 K for all interfaces in Section 3.2. \autoref{fig:smd_pulling_force} plots the resulting force--separation distance curves, and shows the $W_\mathrm{ad}$ values in the inset figure calculated using \autoref{Jarzynski} (Table S1 summarizes the exact values of peak forces and $W_\mathrm{ad}$). The force refers to the pulling force applied to the Cu layers, and the peak force denotes the maximum pulling force experienced immediately before the interface failure. 
Note that the nonequilibrium works used in \autoref{Jarzynski} are not the area under the force--separation distance curves, but area under the force--spring displacement curves. The separation distance refers to the actual separated distance of the pulled group relative to its initial position, while the spring displacement represents the cumulative displacement of the virtual spring tether according to the constant-velocity pulling. Due to the finite spring constant and system resistance during deformation, the actual separation distance of the atoms typically lags behind the pulled distance of the virtual spring. 
\autoref{fig:SMD_config_Ta2N}, Figure S6, and Figure S7 provide structural snapshots in three specific simulation stages: peak force stage ($t = t_\text{peak}$), 0.5 ns after peak force stage ($t = t_\text{peak}+0.5\,\text{ns}$), and at the end of the simulation ($t = t_\text{end}$). 

Relaxed interfaces are consistently marked by the lowest peak forces across all Ta stoichiometries, reflecting their minimal intermixing. Specifically, the relaxed Cu/a-TaN interface demonstrates particularly weak adhesion, characterized by almost complete adhesive failure at the interface leaving minimal residual Cu atoms on the substrate, as observed in Figure S6c. As the Ta content increases from TaN to Ta$_4$N, relaxed interfaces display improved peak forces and start to shift toward cohesive failure modes, evidenced by the increase of residual Cu atoms after delamination (\autoref{fig:SMD_config_Ta2N}c and Figure S7c).

The annealed and deposited interfaces generally show higher peak forces compared to the relaxed interfaces. Similar to relaxed cases, annealed interfaces also demonstrate gradual improvements in peak force at elevated Ta stoichiometries, although the magnitude of these increments is less pronounced. In contrast, deposited interfaces achieve peak force saturation already at the Ta$_2$N composition, with virtually no further enhancement when moving from Ta$_2$N to Ta$_4$N. 

In addition to the trends observed for peak forces, analyzing $W_{\mathrm{ad}}$ reveals additional insights into the mechanical response. For the Cu/a-TaN interfaces, all three conditions present sharp force drops to zero immediately after the peak force, indicative of brittle fracture. Structural snapshots (Figure S6b,e,h) confirm rapid and clean separation at the interface, resulting in relatively low $W_{\mathrm{ad}}$ values.

For the Cu/a-Ta$_2$N and Cu/a-Ta$_4$N interfaces, annealed cases demonstrate enhanced ductility, with structural snapshots (\autoref{fig:SMD_config_Ta2N}d and Figure S7d) revealing partial dislocation nucleation. This plastic deformation significantly contributes to high $W_{\mathrm{ad}}$ values. However, it is important to note that nucleation of partial dislocations alone does not 
imply significant ductility. For instance, partial dislocations are observed across all relaxed interfaces at peak force stages (\autoref{fig:SMD_config_Ta2N}a, Figure S6a, and Figure S7a), but the actual ductility varies considerably with Ta content. This can be understood through the Rice--Thomson fracture criterion\cite{Rice01011974}, which states that ductile fracture occurs only if the energy required to nucleate and propagate these dislocations remains consistently lower than the energy required for cleavage (interface bond breaking in the present context). Thus, ductility increases progressively with Ta content in relaxed interfaces. 

Finally, deposited interfaces show predominantly confined deformation within the intermixing region, without partial dislocation nucleation (\autoref{fig:SMD_config_Ta2N}g and Figure S7g). Despite limited activation of partial dislocations, deposited interfaces yield high $W_{\mathrm{ad}}$ values, primarily due to their high peak forces. The detailed relationship between interfacial structure, stress/strain distribution, and deformation behavior across various interfaces will be examined in the following section.

\subsubsection{Stress/strain distribution and deformation behavior}
\begin{figure}[t!]
    \centering
    \includegraphics{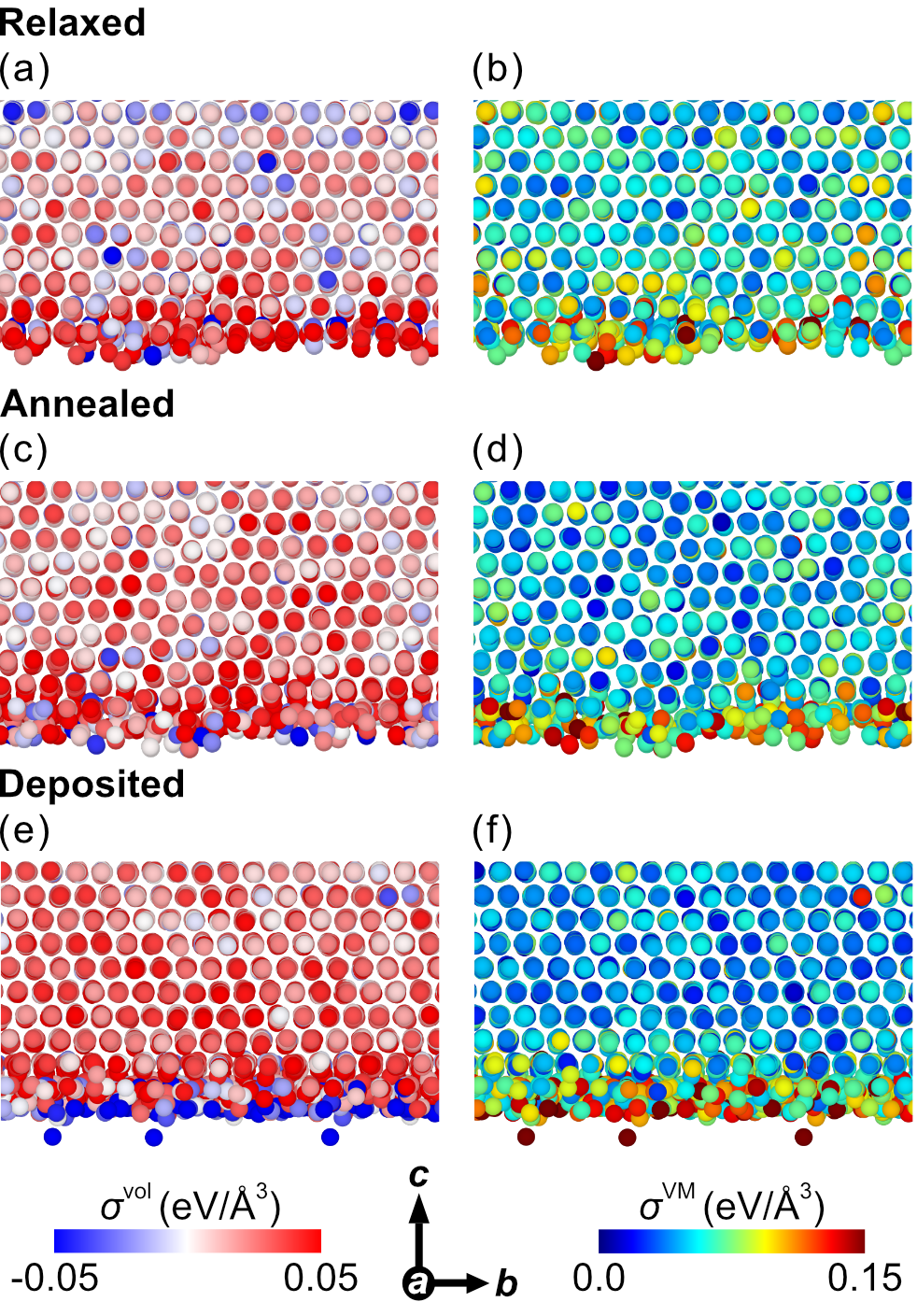}
    \caption{Side-view snapshots of volumetric (left) and von Mises (right) atomic stresses at Cu/a-Ta$_2$N interfaces constructed by (a,b) relaxation, (c,d) annealing, and (e,f) deposition, shown at an intermediate SMD stage ($t=0.5\,\text{ns}$). Ta and N atoms are omitted for clarity.}
    \label{fig:Ta2N_stress}
\end{figure}

\begin{figure*}[t!]
    \centering
    \includegraphics{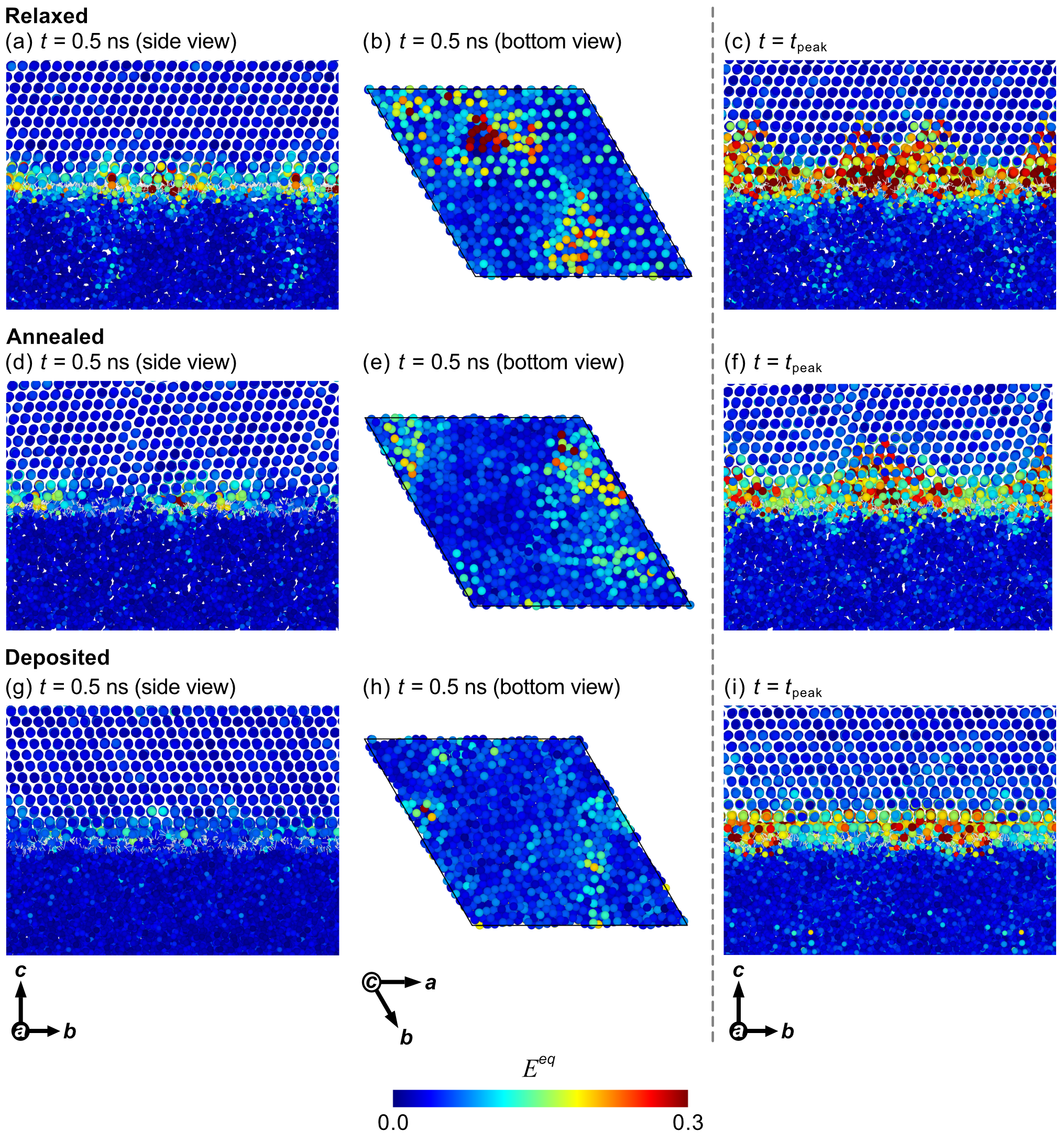}
    \caption{
    Snapshots of the Cu/a-Ta$_2$N interfaces during SMD simulations, colored by atomic equivalent strain computed relative to the initial structure at the start of the SMD simulation. Top, middle, and bottom panels correspond to relaxed, annealed, and deposited interfaces, respectively. (a,d,g) Side views of at $t=0.5\,\text{ns}$. (b,e,h) Bottom views at $t=0.5\,\text{ns}$, showing Cu atoms only (Ta and N omitted) to highlight strain distribution in the Cu layer directly contacting Ta$_2$N. (c,f,i) Side views at $t = t_\text{peak}$. White bonds indicate Cu--Ta and Cu--N bonds.
    }
    \label{fig:strain_unified_Ta2N}
\end{figure*}

To understand the connection between the varied interfacial structures and the observed fracture behavior, we perform a detailed analysis of atomic stresses computed using the virial theorem:

\vspace{0.0em}
\begin{equation}
\sigma_{i,\alpha\beta}
= -\frac{1}{V_i}
\left[
m_i v_{i\alpha} v_{i\beta}
+ \frac{1}{2}\sum_{j\neq{}i} f_{ij\alpha}\,r_{ij\beta}
\right]
\label{equ:stress}
\end{equation}

In \autoref{equ:stress}, $\sigma_{i,\alpha\beta}$ represents the components of the atomic stress tensor of atom $i$, where the indices $\alpha$ and $\beta$ correspond to Cartesian directions. Here, $m_i$ is the mass of atom $i$; $v_{i\alpha}$ and $v_{i\beta}$ are the velocity components; and $f_{ij\alpha}$ and $r_{ij\beta}$ are the pairwise force and displacement components between atoms $i$ and $j$, respectively. The atomic volume $V_i$ is evaluated using Voronoi tessellation. From this stress tensor, the atomic volumetric stress ($\sigma_i^{\text{vol}}$) is calculated as the average of the normal stress components:

\begin{equation}
\sigma_i^{\text{vol}} = \frac{\sigma_{i,xx} + \sigma_{i,yy} + \sigma_{i,zz}}{3}
\label{equ:volumetric}
\end{equation}

Similarly, we calculate the atomic von Mises stress ($\sigma_i^{\text{VM}}$) as:

\vspace{-0.0em}
\begin{multline}
\sigma_i^{\text{VM}} = \frac{1}{\sqrt{2}} \bigl[ (\sigma_{i,xx}-\sigma_{i,yy})^2 + (\sigma_{i,yy}-\sigma_{i,zz})^2 \\ + (\sigma_{i,zz}-\sigma_{i,xx})^2 
+ 6(\sigma_{i,xy}^2+\sigma_{i,yz}^2+\sigma_{i,zx}^2) \bigr]^{1/2}
\label{equ:vonMises}
\end{multline}

\autoref{fig:Ta2N_stress}, Figure S8, and Figure S9 show snapshots of each interface (Ta and N atoms omitted for clarity), colored by $\sigma^{\text{vol}}$ and $\sigma^{\text{VM}}$. Across all Ta stoichiometries, a progression in the volumetric stress distribution emerges from relaxed through annealed to deposited interfaces. Deposited interfaces prominently contain atoms under both tensile (positive) and compressive (negative) volumetric stress within the intermixing region. These compressively stressed atoms originate from depressions in the a-Ta$_x$N layer, resulting in pre-compressed bonds. During pulling, these compressions must first be relieved and reversed to tension, significantly enhancing interfacial robustness. This increasingly robust intermixed structure enables progressively greater stress transfer into the bulk Cu region, a capability that steadily diminishes toward relaxed interfaces, where volumetric stress remains primarily localized at the interface.

A gradual transition is also evident in von Mises stress distributions: from relaxed to annealed to deposited interfaces, shear stress concentrates within the intermixing region due to the increasingly dense and stiff Cu--Ta bonding network. In relaxed interfaces, the relatively soft interface easily yields under shear, limiting its capacity to sustain a high von Mises stress and thus allowing shear stress to propagate deeper into the bulk Cu. As the interface structure transitions from annealed to deposited, the enhanced stiffness of the intermixing region incrementally confines the shear stress. For the annealed case, this stiffness enables the transfer of high loads that drive bulk plasticity. For the deposited case, however, the confinement becomes so effective that it leads to a clear reduction of the bulk region's von Mises stress, thereby suppressing dislocation activity. Atomic stress histograms for bulk region atoms (Figure S10 and Figure S11) quantitatively confirm the aforementioned trends, clearly illustrating a progressive increase in volumetric stress intensity and a decrease in von Mises stress intensity within the Cu bulk region from relaxed to annealed to deposited interfaces.

\begin{figure*}[t!]
    \centering
    \includegraphics{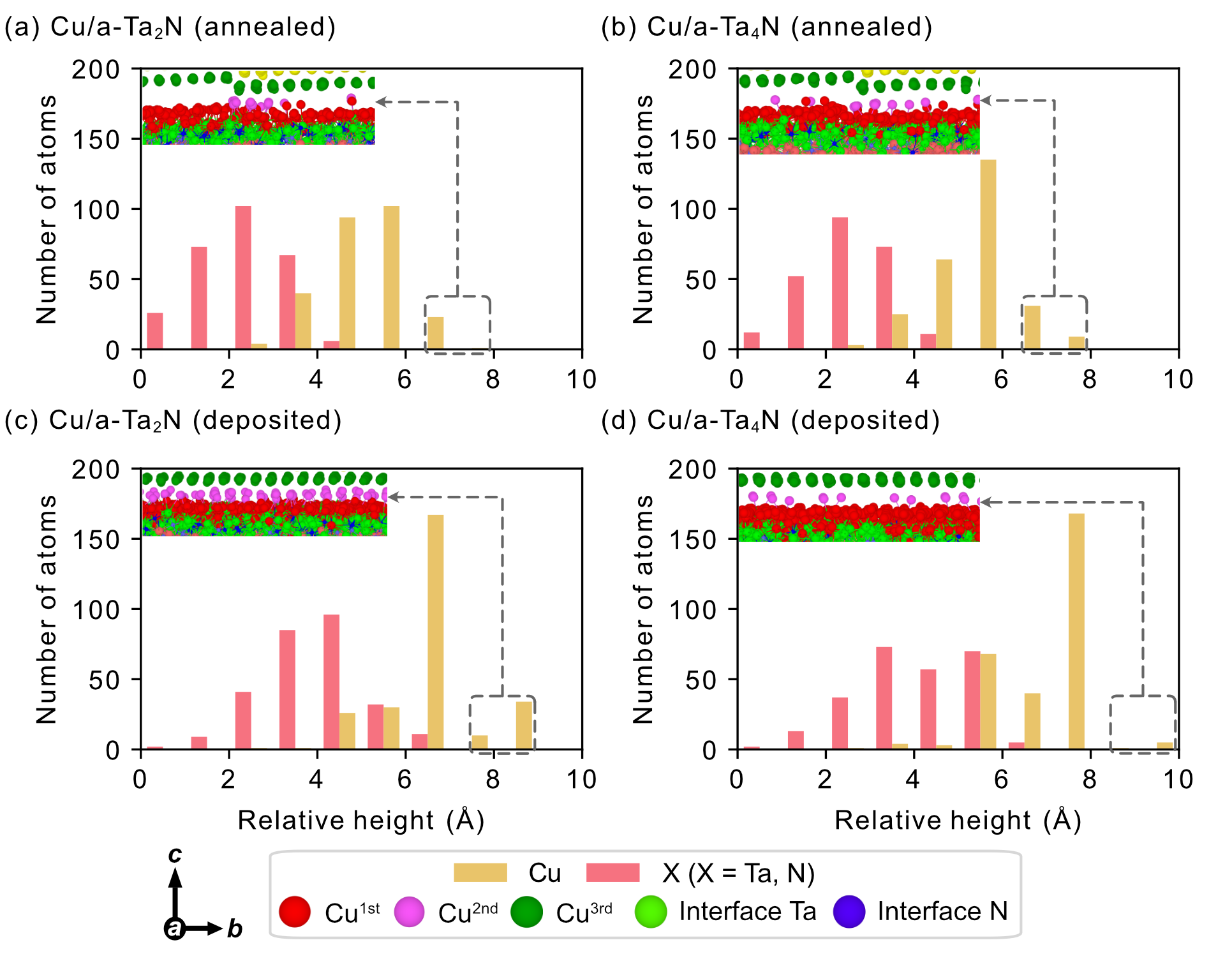} 
    \caption{Number of Cu and X (X = Ta, N) atoms participating in Cu--X bonds as a function of relative height for each interface. Gray dashed boxes indicate Cu atoms located in the second Cu layer. Insets show side views of the corresponding interfaces, where arrows mark the positions of Cu$^{2\mathrm{nd}}$ atoms included in the atom counting (Cu$^{2\mathrm{nd}}$ atoms bonded only to Cu are omitted).}
    \label{fig:atom_num}
\end{figure*}

To understand the relationship between stress distributions and plastic deformation initiation and progression, we also map atomic equivalent strains at specific simulation times ($t = 0.5$ ns and $t = t_\text{peak}$), as shown in \autoref{fig:strain_unified_Ta2N}, Figure S12, and Figure S13. The equivalent strain of atom $i$ ($E_i^{\text{eq}}$) is calculated using the Green--Lagrangian strain tensor components of atom $i$  ($E_{i,\alpha\beta}$):

\vspace{-\baselineskip}
\begin{multline}
E_i^{\text{eq}} = [\frac{1}{6}\{(E_{i,xx}-E_{i,yy})^2+(E_{i,yy}-E_{i,zz})^2 \\ +(E_{i,zz}-E_{i,xx})^2\} + E_{i,xy}^2 + E_{i,yz}^2 + E_{i,zx}^2 ]^{1/2}
\label{equ:strain}
\end{multline}

Distinct differences in strain localization among interface types are observed during the loading stage ($t = 0.5$ ns). Relaxed interfaces of Cu/a-Ta$_2$N and Cu/a-Ta$_4$N manifest pronounced and heterogeneous interfacial strain distributions, characterized by localized regions of intense strain, shown in bottom-view snapshots (\autoref{fig:strain_unified_Ta2N}b and Figure S13b). However, for Cu/a-TaN, relaxed interface shows broadly distributed high-strain regions (Figure S12b), indicative of weak adhesion and direct delamination at the Cu/a-TaN boundary. In contrast, deposited interfaces maintain a homogeneous and reduced strain distribution, reflecting their structural robustness, while annealed interfaces display moderate levels of strain localization.

At the peak force stage ($t = t_\mathrm{peak}$), the impact of these strain differences becomes pronounced. Relaxed interfaces, already marked by significant early bulk von Mises stress, readily facilitate the nucleation of partial dislocations, permitting plastic deformation to extend substantially into the bulk Cu. Deposited interfaces, by effectively confining both shear stress and deformation within the intermixing region, minimize bulk plasticity. Annealed interfaces continue to demonstrate intermediate characteristics, exhibiting moderate strain localization at the interface while still showing partial dislocation nucleation into the bulk region. 

A similar phenomenon has been reported in amorphous intergranular films, where the disordered interface locally accommodates plastic deformation, redistributing strain and hindering dislocation propagation under dislocation-induced stress \cite{10.1016/j.actamat.2015.02.012}. Our intermixing layers similarly serve as effective strain sinks preventing extended defect propagation into the crystalline Cu. Moreover, this strain localization provides a plausible explanation for the observed marginal ductility in the deposited Cu/a-Ta$_4$N interfaces without dislocation activity, described earlier (Section 3.3.1). Overall, these observations demonstrate that the intermixing not only enhances adhesion but also significantly influences the spatial distribution and progression of plastic deformation.

\subsubsection{Effect of vertical distribution of Ta}

In Section 3.3.1, we observe an intriguing result: the peak forces and $W_{\mathrm{ad}}$ for the deposited interface of Cu/a-Ta$_2$N and Cu/a-Ta$_4$N yield nearly identical values. This saturation suggests that the system's strength is no longer limited by direct interfacial adhesion but by the cohesive strength of the intermixing region itself. Therefore, to understand this behavior, we focus on the atomic-scale interactions within the first few Cu layers where fracture is observed.

For the Cu/a-Ta$_2$N and Cu/a-Ta$_4$N cases, fracture does not occur directly at the interface, leaving the plenty of residual Cu atoms on the Ta$_x$N surface after the delamination (\autoref{fig:SMD_config_Ta2N}c,f,i
and Figure S7c,f,i). Therefore, it is expected that the interactions between the second-layer Cu atoms (Cu$^{2\mathrm{nd}}$) and first-layer Cu atoms (Cu$^{\mathrm{1st}}$), rather than between Cu$^{\mathrm{1st}}$ and Ta atoms, become crucial as Ta content increases, influencing the cohesive failure (see Figure S14 for definition of Cu layers).

\autoref{fig:atom_num} shows the count of atoms that form Cu--Ta or Cu--N bonds as a function of vertical height. Note that the inset omits Cu$^{\mathrm{2nd}}$ atoms that do not form the Cu--Ta or Cu--N bonds to clearly visualize the vertical distribution of Cu$^{\mathrm{2nd}}$ atoms directly involved in interfacial bonding. Remarkably, the deposited Cu/a-Ta$_2$N interface displays greater Ta diffusion up to the Cu$^{\mathrm{1st}}$ layer, whereas the other three interfaces confine Ta atoms closer to the original interface boundary. This Ta diffusion is critical, as it enables Cu$^{\mathrm{2nd}}$ atoms to simultaneously form both Cu$^{\mathrm{2nd}}$--Ta bonds with incorporated Ta atoms and Cu$^{\mathrm{2nd}}$--Cu$^{\mathrm{1st}}$ bonds. Given the known covalent character of Cu--Ta bonding at Cu/Ta$_2$N interfaces \cite{10.1039/c8cp01839a}, these additional bonds reinforce the cohesive strength within the first two Cu layers.

\begin{figure}[t!]
    \centering
    \includegraphics{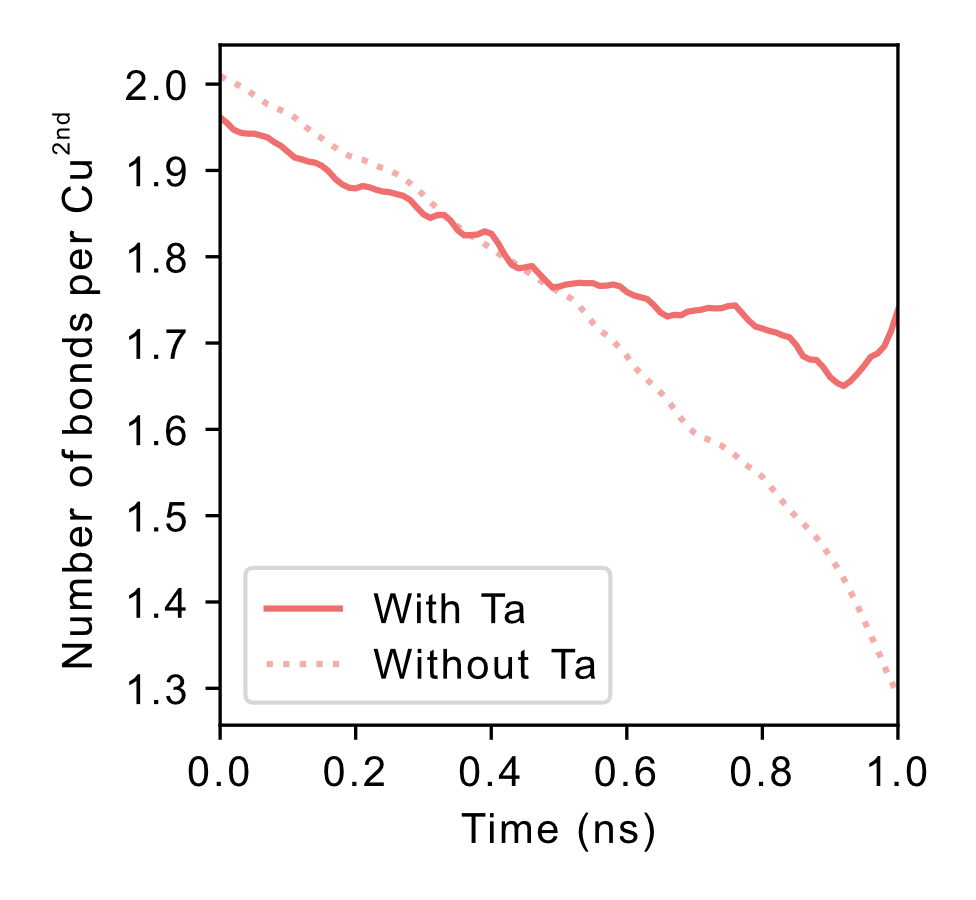}
    \caption{Evolution of the average numbers of bonds per Cu$^{2\mathrm{nd}}$ atom during SMD simulation of the deposited Cu/a-Ta$_2$N interface. Solid line: Cu$^{2\mathrm{nd}}$ atoms forming Cu--Ta bonds (including both Cu$^{2\mathrm{nd}}$--Cu$^{1\mathrm{st}}$ and Cu$^{2\mathrm{nd}}$--Ta bonds). Dotted line: Cu$^{2\mathrm{nd}}$ atoms without Cu--Ta bonding (only Cu$^{2\mathrm{nd}}$--Cu$^{1\mathrm{st}}$ bonds are counted).}
    \label{fig:bond_num_time}
\end{figure}

\begin{figure*}[ht]
  \centering
  \includegraphics{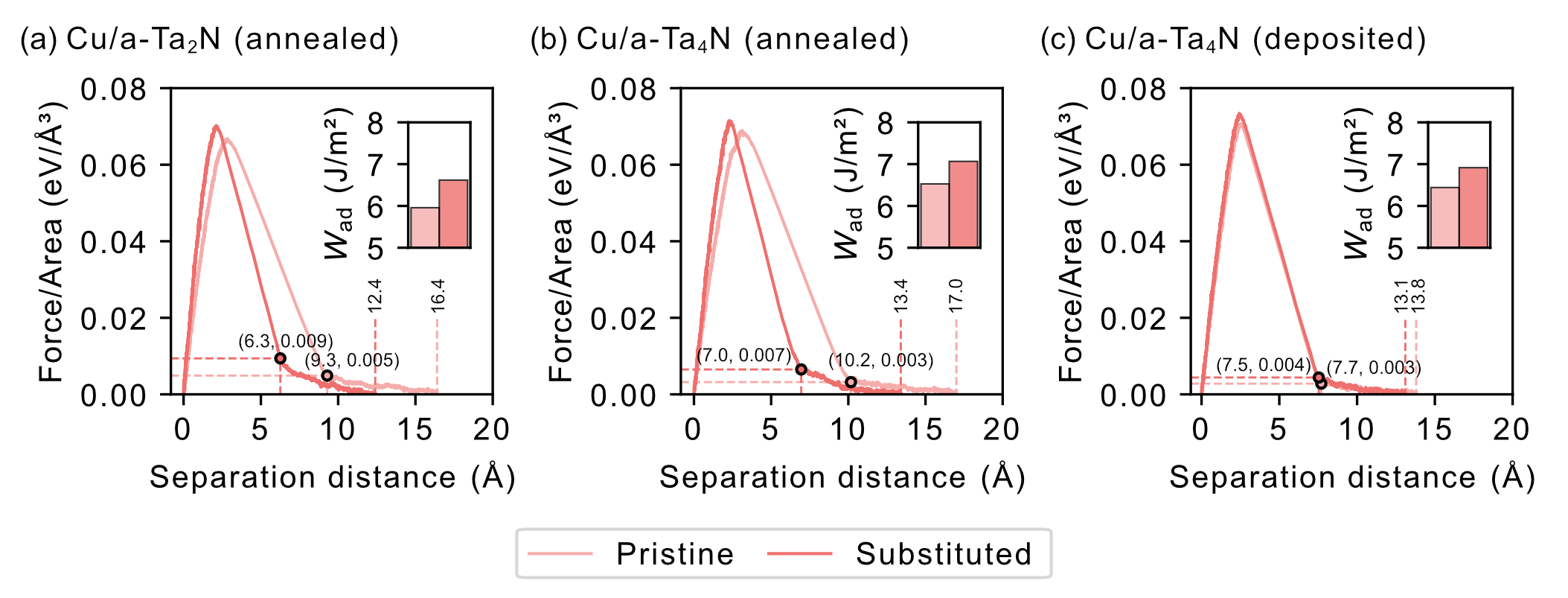}
  \caption{Comparison of pulling force evolution during SMD simulations for pristine (light red) and substituted (red) Cu/a-Ta$_x$N interfaces: (a) Cu/a-Ta$_2$N (annealed), (b) Cu/a-Ta$_4$N (annealed), (c) Cu/a-Ta$_4$N (deposited). Circular markers and their coordinate labels denote the start of the tail region, while dashed vertical lines and corresponding numbers indicate separation distances where the PMF converges within 10$^{-4}~\mathrm{eV/}$\AA$^2$. Insets show the corresponding $W_\mathrm{ad}$ obtained from each simulation.}
  \label{fig:smd_pulling_force_subs}
\end{figure*}

\autoref{fig:bond_num_time} presents the evolution of the bond numbers of Cu$^{2\mathrm{nd}}$ atoms during the SMD process, for the deposited Cu/a-Ta$_2$N interface. Each curve represents the average bond count per atom within its respective group: Cu$^{2\mathrm{nd}}$ atoms whether they have any Ta neighbor atom (solid line), or not (dotted line). It is confirmed that the number of bonds in the Cu$^{\mathrm{2nd}}$ layer decreases more gradually when these atoms maintain both Cu--Cu and Cu--Ta interactions.
To further verify the contribution of incorporated Ta atoms, about 10\% of Cu$^{\mathrm{1st}}$ atoms are randomly chosen and substituted with Ta atoms in the annealed Cu/a-Ta$_2$N, the annealed Cu/a-Ta$_4$N, and the deposited Cu/a-Ta$_4$N interfaces (Figure S15). This substitution results in a Cu$^{\mathrm{1st}}$--Ta bound count similar to that of the deposited Cu/a-Ta$_2$N interface. After additional equilibration at 300 K, SMD simulations are performed for the substituted interfaces, with the resulting force--separation distance curves shown in \autoref{fig:smd_pulling_force_subs} with the original data of each interface for comparison. The substitution of Ta atoms consistently enhances the interfacial strength, evidenced by the higher peak forces in all tested cases. Furthermore, the force--separation curves for the substituted interfaces exhibit a steeper initial slope and reach their peak force at a smaller separation distance, indicating a improvement in interfacial stiffness.

This enhanced stiffness is coupled with a notable change in the post-peak fracture behavior, particularly for the annealed interfaces (\autoref{fig:smd_pulling_force_subs}a,b). The onset of the plastic deformation regime, marked as the start of the tail region, occurs at a higher sustained force in the substituted cases compared to the original ones (e.g., 0.009 eV/\AA$^3$ vs. 0.005 eV/\AA$^3$ for the annealed Cu/a-Ta$_2$N interface). This demonstrates that after the initiation of failure, the substituted interface maintains a greater load-bearing capacity during plastic deformation, signifying a more robust ductile response. In contrast, for the deposited interface (\autoref{fig:smd_pulling_force_subs}c), the substitution primarily increases the peak force with minimal alteration to the subsequent fracture characteristics; the tail regions for both original and substituted systems commence at similar force levels and follow a nearly identical decay path.

These observations suggest that the effect of Ta substitution is mechanism-dependent. For annealed interfaces, it appears to enhance ductility by promoting dislocation-mediated plasticity under higher stress. Conversely, for the deposited interface, where deformation is inherently confined to the intermixing region without significant dislocation activation, the substitution primarily serves to increase the cohesive strength. Consistent with these mechanical improvements, the $W_{\mathrm{ad}}$ increases for all substituted cases, as shown in the inset of \autoref{fig:smd_pulling_force_subs} and Table S2.

Overall, the results emphasize the distinct impact of the targeted Ta substitution depending on the interface construction and the associated deformation mechanisms. Moreover, these findings provide additional insights and strategies for approaches aimed at developing liner-free barrier materials through atomic-scale engineering, such as strategic metal doping at the interface to improve adhesion properties. For instance, recent computational study by Ref.~\citenum{10.1039/d1sc04708f} has investigated Ru incorporation into TaN to promote Cu wetting and adhesion in interconnect applications. Their work emphasizes that carefully tuning metal dopants at the interface significantly influences adhesion properties by controlling interfacial bonding interactions and local atomic environments. Our analysis further highlights the importance of targeted vertical distribution of doped elements, underscoring that controlled Ta incorporation into the Cu layers provides adhesion improvements beyond simply increasing the overall dopant content.

\section{Conclusion}

In summary, the MLIP has been employed to systematically investigate adhesion and deformation behaviors at Cu/a-Ta$_x$N interfaces under varying stoichiometries and interface construction conditions. The developed MLIP accurately captures key atomic interactions, enabling comprehensive SMD simulations that elucidate the mechanisms underlying interface strength and failure dynamics.

Our results indicate that higher Ta content enhances interfacial adhesion by fostering dense and robust Cu--Ta bonding networks. In particular, annealed and deposited interfaces exhibit pronounced atomic intermixing, resulting in predominantly cohesive failures within the Cu layer rather than adhesive interfacial fractures. Among them, annealed interfaces achieve enhanced ductility and higher overall work of adhesion through substantial plastic deformation that extends into bulk Cu layers, while deposited interfaces yield higher peak strengths but limited ductility due to strong but more localized bonding networks. Notably, increasing Ta content alone reaches a saturation point in peak force, constraining further improvement.

We further identify targeted Ta incorporation into the Cu layers as a means to strengthen cohesive interactions beyond this saturation limit, highlighting a potential strategy for optimizing interface performance via precise atomic-scale engineering. Overall, our findings emphasize the critical interplay between interface stoichiometry, processing, and mechanical response, providing practical guidance for the design of liner-free, high-performance interconnect systems.

\begin{acknowledgement}

This work was supported by Samsung Electronics Co., Ltd(IO201214-08143-01) and the National Research Foundation of Korea (NRF) funded by the Korean government (MSIP) (2022M3C1C8093916). The computations were carried out at Korea Institute of Science and Technology Information (KISTI) National Supercomputing Center (KSC-2024-CRE-0215).

\end{acknowledgement}

\begin{suppinfo}

\begin{itemize}
  \item Supporting Information: peak forces and work of adhesion values for Cu/a-Ta$_x$N interfaces, comparison of the DFT and MLIP energies and force components along the MLIP MD trajectory, SMD parameter convergence test, side view of generated Cu/a-TaN and Cu/a-Ta$_4$N interface structures, surface meshes, snapshots from SMD simulations of Cu/a-TaN and Cu/a-Ta$_4$N interfaces, volumetric and von Mises atomic stresses of Cu/a-TaN and Cu/a-Ta$_4$N interfaces, distribution of volumetric and von Mises atomic stresses, atomic equivalent strain of Cu/a-TaN and Cu/a-Ta$_4$N interfaces, layer definition of interfaces, substituted Cu/a-Ta$_4$N interface structure
\end{itemize}

\end{suppinfo}

\bibliography{achemso-demo}

\providecommand{\latin}[1]{#1}
\makeatletter
\providecommand{\doi}
  {\begingroup\let\do\@makeother\dospecials
  \catcode`\{=1 \catcode`\}=2 \doi@aux}
\providecommand{\doi@aux}[1]{\endgroup\texttt{#1}}
\makeatother
\providecommand*\mcitethebibliography{\thebibliography}
\csname @ifundefined\endcsname{endmcitethebibliography}  {\let\endmcitethebibliography\endthebibliography}{}
\begin{mcitethebibliography}{74}
\providecommand*\natexlab[1]{#1}
\providecommand*\mciteSetBstSublistMode[1]{}
\providecommand*\mciteSetBstMaxWidthForm[2]{}
\providecommand*\mciteBstWouldAddEndPuncttrue
  {\def\EndOfBibitem{\unskip.}}
\providecommand*\mciteBstWouldAddEndPunctfalse
  {\let\EndOfBibitem\relax}
\providecommand*\mciteSetBstMidEndSepPunct[3]{}
\providecommand*\mciteSetBstSublistLabelBeginEnd[3]{}
\providecommand*\EndOfBibitem{}
\mciteSetBstSublistMode{f}
\mciteSetBstMaxWidthForm{subitem}{(\alph{mcitesubitemcount})}
\mciteSetBstSublistLabelBeginEnd
  {\mcitemaxwidthsubitemform\space}
  {\relax}
  {\relax}

\bibitem[Edelstein(2017)]{8268387}
Edelstein,~D.~C. 20 Years of Cu BEOL in manufacturing, and its future prospects. 2017 IEEE International Electron Devices Meeting (IEDM). 2017; pp 14.1.1--14.1.4\relax
\mciteBstWouldAddEndPuncttrue
\mciteSetBstMidEndSepPunct{\mcitedefaultmidpunct}
{\mcitedefaultendpunct}{\mcitedefaultseppunct}\relax
\EndOfBibitem
\bibitem[Corn \latin{et~al.}(1988)Corn, Falconer, and Czanderna]{10.1116/1.575620}
Corn,~S.~H.; Falconer,~J.~L.; Czanderna,~A.~W. {The copper–silicon interface: Composition and interdiffusion}. \emph{J. Vac. Sci. Technol. A} \textbf{1988}, \emph{6}, 1012--1016\relax
\mciteBstWouldAddEndPuncttrue
\mciteSetBstMidEndSepPunct{\mcitedefaultmidpunct}
{\mcitedefaultendpunct}{\mcitedefaultseppunct}\relax
\EndOfBibitem
\bibitem[Chang(1990)]{10.1063/1.345194}
Chang,~C.-A. {Formation of copper silicides from Cu(100)/Si(100) and Cu(111)/Si(111) structures}. \emph{J. Appl. Phys.} \textbf{1990}, \emph{67}, 566--569\relax
\mciteBstWouldAddEndPuncttrue
\mciteSetBstMidEndSepPunct{\mcitedefaultmidpunct}
{\mcitedefaultendpunct}{\mcitedefaultseppunct}\relax
\EndOfBibitem
\bibitem[Min(1996)]{10.1116/1.588818}
Min,~K.-H. {Comparative study of tantalum and tantalum nitrides (Ta$_2$N and TaN) as a diffusion barrier for Cu metallization}. \emph{J. Vac. Sci. Technol. B} \textbf{1996}, \emph{14}, 3263\relax
\mciteBstWouldAddEndPuncttrue
\mciteSetBstMidEndSepPunct{\mcitedefaultmidpunct}
{\mcitedefaultendpunct}{\mcitedefaultseppunct}\relax
\EndOfBibitem
\bibitem[Wang \latin{et~al.}(2002)Wang, Chen, Lu, Hsiung, Hsieh, and Yew]{10.1116/1.1495906}
Wang,~J.~H.; Chen,~L.~J.; Lu,~Z.~C.; Hsiung,~C.~S.; Hsieh,~W.~Y.; Yew,~T.~R. {Ta and Ta–N diffusion barriers sputtered with various N$_2$/Ar ratios for Cu metallization}. \emph{J. Vac. Sci. Technol. B} \textbf{2002}, \emph{20}, 1522--1526\relax
\mciteBstWouldAddEndPuncttrue
\mciteSetBstMidEndSepPunct{\mcitedefaultmidpunct}
{\mcitedefaultendpunct}{\mcitedefaultseppunct}\relax
\EndOfBibitem
\bibitem[Kim \latin{et~al.}(2008)Kim, Lee, and Lee]{10.1143/jjap.47.6953}
Kim,~S.-M.; Lee,~G.-R.; Lee,~J.-J. {Effect of Film Microstructure on Diffusion Barrier Properties of TaN$_x$ Films in Cu Metallization}. \emph{Jpn. J. Appl. Phys.} \textbf{2008}, \emph{47}, 6953--6955\relax
\mciteBstWouldAddEndPuncttrue
\mciteSetBstMidEndSepPunct{\mcitedefaultmidpunct}
{\mcitedefaultendpunct}{\mcitedefaultseppunct}\relax
\EndOfBibitem
\bibitem[Kaloyeros and Eisenbraun(2000)Kaloyeros, and Eisenbraun]{10.1146/annurev.matsci.30.1.363}
Kaloyeros,~A.~E.; Eisenbraun,~E. {Ultrathin Diffusion Barriers/Liners for Gigascale Copper Metallization}. \emph{Annu. Rev. Mater. Sci.} \textbf{2000}, \emph{30}, 363--385\relax
\mciteBstWouldAddEndPuncttrue
\mciteSetBstMidEndSepPunct{\mcitedefaultmidpunct}
{\mcitedefaultendpunct}{\mcitedefaultseppunct}\relax
\EndOfBibitem
\bibitem[Holloway \latin{et~al.}(1992)Holloway, Fryer, Cabral, Harper, Bailey, and Kelleher]{10.1063/1.350566}
Holloway,~K.; Fryer,~P.~M.; Cabral,~C.; Harper,~J. M.~E.; Bailey,~P.~J.; Kelleher,~K.~H. {Tantalum as a diffusion barrier between copper and silicon: Failure mechanism and effect of nitrogen additions}. \emph{J. Appl. Phys.} \textbf{1992}, \emph{71}, 5433--5444\relax
\mciteBstWouldAddEndPuncttrue
\mciteSetBstMidEndSepPunct{\mcitedefaultmidpunct}
{\mcitedefaultendpunct}{\mcitedefaultseppunct}\relax
\EndOfBibitem
\bibitem[Edelstein \latin{et~al.}(2001)Edelstein, Uzoh, Cabral, DeHaven, Buchwalter, Simon, Cooney, Malhotra, Klaus, Rathore, Agarwala, and Nguyen]{10.1109/iitc.2001.930001}
Edelstein,~D.; Uzoh,~C.; Cabral,~C.; DeHaven,~P.; Buchwalter,~P.; Simon,~A.; Cooney,~E.; Malhotra,~S.; Klaus,~D.; Rathore,~H.; Agarwala,~B.; Nguyen,~D. {A high performance liner for copper damascene interconnects}. Proceedings of the IEEE 2001 International Interconnect Technology Conference (Cat. No.01EX461). 2001; pp 9--11\relax
\mciteBstWouldAddEndPuncttrue
\mciteSetBstMidEndSepPunct{\mcitedefaultmidpunct}
{\mcitedefaultendpunct}{\mcitedefaultseppunct}\relax
\EndOfBibitem
\bibitem[He \latin{et~al.}(2013)He, Zhang, Nogami, Lin, Kelly, Kim, Spooner, Edelstein, and Zhao]{10.1149/2.009312jes}
He,~M.; Zhang,~X.; Nogami,~T.; Lin,~X.; Kelly,~J.; Kim,~H.; Spooner,~T.; Edelstein,~D.; Zhao,~L. {Mechanism of Co Liner as Enhancement Layer for Cu Interconnect Gap-Fill}. \emph{J. Electrochem. Soc.} \textbf{2013}, \emph{160}, D3040--D3044\relax
\mciteBstWouldAddEndPuncttrue
\mciteSetBstMidEndSepPunct{\mcitedefaultmidpunct}
{\mcitedefaultendpunct}{\mcitedefaultseppunct}\relax
\EndOfBibitem
\bibitem[Yang \latin{et~al.}(2006)Yang, Spooner, Ponoth, Chanda, Simon, Lavoie, Lane, Hu, Liniger, Gignac, Shaw, Cohen, McFeely, and Edelstein]{10.1109/iitc.2006.1648684}
Yang,~C.-C.; Spooner,~T.; Ponoth,~S.; Chanda,~K.; Simon,~A.; Lavoie,~C.; Lane,~M.; Hu,~C.-K.; Liniger,~E.; Gignac,~L.; Shaw,~T.; Cohen,~S.; McFeely,~F.; Edelstein,~D. {Physical, Electrical, and Reliability Characterization of Ru for Cu Interconnects}. 2006 International Interconnect Technology Conference. 2006; pp 187--190\relax
\mciteBstWouldAddEndPuncttrue
\mciteSetBstMidEndSepPunct{\mcitedefaultmidpunct}
{\mcitedefaultendpunct}{\mcitedefaultseppunct}\relax
\EndOfBibitem
\bibitem[Chang \latin{et~al.}(2010)Chang, Pan, and Chen]{10.1149/1.3267881}
Chang,~C.-C.; Pan,~F.-M.; Chen,~C.-W. {Effect of Surface Reduction Treatments of Plasma-Enhanced Atomic Layer Chemical Vapor Deposited TaN$_x$ on Adhesion with Copper}. \emph{J. Electrochem. Soc.} \textbf{2010}, \emph{157}, G62\relax
\mciteBstWouldAddEndPuncttrue
\mciteSetBstMidEndSepPunct{\mcitedefaultmidpunct}
{\mcitedefaultendpunct}{\mcitedefaultseppunct}\relax
\EndOfBibitem
\bibitem[Furuya \latin{et~al.}(2005)Furuya, Tsuda, and Ogawa]{10.1116/1.1926289}
Furuya,~A.; Tsuda,~H.; Ogawa,~S. Ta-rich atomic layer deposition TaN adhesion layer for Cu interconnects by means of plasma-enhanced atomic layer deposition. \emph{J. Vac. Sci. Technol. B} \textbf{2005}, \emph{23}, 979--983\relax
\mciteBstWouldAddEndPuncttrue
\mciteSetBstMidEndSepPunct{\mcitedefaultmidpunct}
{\mcitedefaultendpunct}{\mcitedefaultseppunct}\relax
\EndOfBibitem
\bibitem[Sekiguchi and Koike(2008)Sekiguchi, and Koike]{10.1143/JJAP.47.1042}
Sekiguchi,~A.; Koike,~J. {Evaluation of Interface Adhesion Strength in Cu/(Ta–x\% N, Ta/TaN)/SiO$_2$/Si by Nanoscratch Test}. \emph{Jpn. J. Appl. Phys.} \textbf{2008}, \emph{47}, 1042\relax
\mciteBstWouldAddEndPuncttrue
\mciteSetBstMidEndSepPunct{\mcitedefaultmidpunct}
{\mcitedefaultendpunct}{\mcitedefaultseppunct}\relax
\EndOfBibitem
\bibitem[Stavrev \latin{et~al.}(1999)Stavrev, Fischer, Praessler, Wenzel, and Drescher]{10.1116/1.581697}
Stavrev,~M.; Fischer,~D.; Praessler,~F.; Wenzel,~C.; Drescher,~K. {Behavior of thin Ta-based films in the Cu/barrier/Si system}. \emph{J. Vac. Sci. Technol. A} \textbf{1999}, \emph{17}, 993--1001\relax
\mciteBstWouldAddEndPuncttrue
\mciteSetBstMidEndSepPunct{\mcitedefaultmidpunct}
{\mcitedefaultendpunct}{\mcitedefaultseppunct}\relax
\EndOfBibitem
\bibitem[Chen and Chen(2000)Chen, and Chen]{10.1063/1.373566}
Chen,~G.~S.; Chen,~S.~T. {Diffusion barrier properties of single- and multilayered quasi-amorphous tantalum nitride thin films against copper penetration}. \emph{J. Appl. Phys.} \textbf{2000}, \emph{87}, 8473--8482\relax
\mciteBstWouldAddEndPuncttrue
\mciteSetBstMidEndSepPunct{\mcitedefaultmidpunct}
{\mcitedefaultendpunct}{\mcitedefaultseppunct}\relax
\EndOfBibitem
\bibitem[Chen \latin{et~al.}(2000)Chen, Chen, Yang, and Lee]{10.1116/1.582166}
Chen,~G.~S.; Chen,~S.~T.; Yang,~L.-C.; Lee,~P.~Y. {Evaluation of single- and multilayered amorphous tantalum nitride thin films as diffusion barriers in copper metallization}. \emph{J. Vac. Sci. Technol. A} \textbf{2000}, \emph{18}, 720--723\relax
\mciteBstWouldAddEndPuncttrue
\mciteSetBstMidEndSepPunct{\mcitedefaultmidpunct}
{\mcitedefaultendpunct}{\mcitedefaultseppunct}\relax
\EndOfBibitem
\bibitem[Jayaram(2022)]{10.1146/annurev-matsci-080819-123640}
Jayaram,~V. {Small-Scale Mechanical Testing}. \emph{Annu. Rev. Mater. Res.} \textbf{2022}, \emph{52}, 473--523\relax
\mciteBstWouldAddEndPuncttrue
\mciteSetBstMidEndSepPunct{\mcitedefaultmidpunct}
{\mcitedefaultendpunct}{\mcitedefaultseppunct}\relax
\EndOfBibitem
\bibitem[Zhao and Lu(2009)Zhao, and Lu]{10.1103/physrevb.79.214104}
Zhao,~Y.; Lu,~G. {First-principles simulations of copper diffusion in tantalum and tantalum nitride}. \emph{Phys. Rev. B} \textbf{2009}, \emph{79}, 214104\relax
\mciteBstWouldAddEndPuncttrue
\mciteSetBstMidEndSepPunct{\mcitedefaultmidpunct}
{\mcitedefaultendpunct}{\mcitedefaultseppunct}\relax
\EndOfBibitem
\bibitem[Wang \latin{et~al.}(2018)Wang, Ma, Li, Jiang, Chen, and Jiang]{10.1039/c8cp01839a}
Wang,~J.; Ma,~A.; Li,~M.; Jiang,~J.; Chen,~J.; Jiang,~Y. {Chemical bonding and Cu diffusion at the Cu/Ta$_2$N interface: a DFT study.} \emph{Phys. Chem. Chem. Phys.} \textbf{2018}, \emph{20}, 13566--13573\relax
\mciteBstWouldAddEndPuncttrue
\mciteSetBstMidEndSepPunct{\mcitedefaultmidpunct}
{\mcitedefaultendpunct}{\mcitedefaultseppunct}\relax
\EndOfBibitem
\bibitem[Hashibon \latin{et~al.}(2007)Hashibon, Elsässer, Mishin, and Gumbsch]{10.1103/physrevb.76.245434}
Hashibon,~A.; Elsässer,~C.; Mishin,~Y.; Gumbsch,~P. {First-principles study of thermodynamical and mechanical stabilities of thin copper film on tantalum}. \emph{Phys. Rev. B} \textbf{2007}, \emph{76}, 245434\relax
\mciteBstWouldAddEndPuncttrue
\mciteSetBstMidEndSepPunct{\mcitedefaultmidpunct}
{\mcitedefaultendpunct}{\mcitedefaultseppunct}\relax
\EndOfBibitem
\bibitem[Iwamoto \latin{et~al.}(2004)Iwamoto, Truong, and Lee]{10.1016/j.tsf.2004.06.176}
Iwamoto,~N.; Truong,~N.; Lee,~E. {New metal layers for integrated circuit manufacture: experimental and modeling studies}. \emph{Thin Solid Films} \textbf{2004}, \emph{469-470}, 431--437\relax
\mciteBstWouldAddEndPuncttrue
\mciteSetBstMidEndSepPunct{\mcitedefaultmidpunct}
{\mcitedefaultendpunct}{\mcitedefaultseppunct}\relax
\EndOfBibitem
\bibitem[Ding \latin{et~al.}(2010)Ding, Deng, Lu, Jiang, Ru, Zhang, and Qu]{10.1063/1.3369443}
Ding,~S.-F.; Deng,~S.-R.; Lu,~H.-S.; Jiang,~Y.-L.; Ru,~G.-P.; Zhang,~D.~W.; Qu,~X.-P. Cu adhesion on tantalum and ruthenium surface: Density functional theory study. \emph{J. Appl. Phys.} \textbf{2010}, \emph{107}, 103534\relax
\mciteBstWouldAddEndPuncttrue
\mciteSetBstMidEndSepPunct{\mcitedefaultmidpunct}
{\mcitedefaultendpunct}{\mcitedefaultseppunct}\relax
\EndOfBibitem
\bibitem[Li \latin{et~al.}(2017)Li, Wu, Luo, Chen, Tay, and Zhu]{10.1063/1.4997677}
Li,~G.; Wu,~H.; Luo,~H.; Chen,~Z.; Tay,~A. A.~O.; Zhu,~W. {Diffusion behavior of Cu/Ta heterogeneous interface under high temperature and high strain: An atomistic investigation}. \emph{AIP Adv.} \textbf{2017}, \emph{7}, 095320\relax
\mciteBstWouldAddEndPuncttrue
\mciteSetBstMidEndSepPunct{\mcitedefaultmidpunct}
{\mcitedefaultendpunct}{\mcitedefaultseppunct}\relax
\EndOfBibitem
\bibitem[Gong and Liu(2003)Gong, and Liu]{10.1063/1.1630353}
Gong,~H.~R.; Liu,~B.~X. Interface stability and solid-state amorphization in an immiscible Cu-Ta system. \emph{Appl. Phys. Lett.} \textbf{2003}, \emph{83}, 4515--4517\relax
\mciteBstWouldAddEndPuncttrue
\mciteSetBstMidEndSepPunct{\mcitedefaultmidpunct}
{\mcitedefaultendpunct}{\mcitedefaultseppunct}\relax
\EndOfBibitem
\bibitem[Yang \latin{et~al.}(2014)Yang, Li, Shi, and Kong]{10.1016/j.commatsci.2014.01.028}
Yang,~G.; Li,~J.; Shi,~Q.; Kong,~L. {Structural and dynamical properties of heterogeneous solid–liquid Ta–Cu interfaces: A molecular dynamics study}. \emph{Comput. Mater. Sci.} \textbf{2014}, \emph{86}, 64--72\relax
\mciteBstWouldAddEndPuncttrue
\mciteSetBstMidEndSepPunct{\mcitedefaultmidpunct}
{\mcitedefaultendpunct}{\mcitedefaultseppunct}\relax
\EndOfBibitem
\bibitem[Yang \latin{et~al.}(2015)Yang, Gao, Li, and Kong]{10.1063/1.4905103}
Yang,~G.; Gao,~X.; Li,~J.; Kong,~L. Orientation dependences of atomic structures in chemically heterogeneous Cu$_{50}$Ta$_{50}$/Ta glass-crystal interfaces. \emph{J. Appl. Phys.} \textbf{2015}, \emph{117}, 015303\relax
\mciteBstWouldAddEndPuncttrue
\mciteSetBstMidEndSepPunct{\mcitedefaultmidpunct}
{\mcitedefaultendpunct}{\mcitedefaultseppunct}\relax
\EndOfBibitem
\bibitem[Lazi\ifmmode~\acute{c}\else \'{c}\fi{} \latin{et~al.}(2010)Lazi\ifmmode~\acute{c}\else \'{c}\fi{}, Klaver, and Thijsse]{PhysRevB.81.045410}
Lazi\ifmmode~\acute{c}\else \'{c}\fi{},~I.; Klaver,~P.; Thijsse,~B. Microstructure of a Cu film grown on bcc Ta (100) by large-scale molecular-dynamics simulations. \emph{Phys. Rev. B} \textbf{2010}, \emph{81}, 045410\relax
\mciteBstWouldAddEndPuncttrue
\mciteSetBstMidEndSepPunct{\mcitedefaultmidpunct}
{\mcitedefaultendpunct}{\mcitedefaultseppunct}\relax
\EndOfBibitem
\bibitem[Klaver and Thijsse(2011)Klaver, and Thijsse]{10.1557/proc-721-j2.3}
Klaver,~P.; Thijsse,~B.~J. {Molecular Dynamics Study of Cu Thin Film Deposition on $\beta$-Ta}. \emph{MRS Online Proc. Libr.} \textbf{2011}, \emph{721}, 23\relax
\mciteBstWouldAddEndPuncttrue
\mciteSetBstMidEndSepPunct{\mcitedefaultmidpunct}
{\mcitedefaultendpunct}{\mcitedefaultseppunct}\relax
\EndOfBibitem
\bibitem[Klaver and Thijsse(2003)Klaver, and Thijsse]{10.1023/b:jcad.0000036802.46424.ee}
Klaver,~T.; Thijsse,~B. {Molecular Dynamics simulations of Cu/Ta and Ta/Cu thin film growth}. \emph{J. Comput.-Aided Mater. Des.} \textbf{2003}, \emph{10}, 61--74\relax
\mciteBstWouldAddEndPuncttrue
\mciteSetBstMidEndSepPunct{\mcitedefaultmidpunct}
{\mcitedefaultendpunct}{\mcitedefaultseppunct}\relax
\EndOfBibitem
\bibitem[Hong and Yang(2012)Hong, and Yang]{10.1143/jjap.51.06ff14}
Hong,~R.-T.; Yang,~J.-Y. {Molecular Dynamics Study on Enhanced Cu Coverage of Trench Filling with Low-Index Ta Surfaces}. \emph{Jpn. J. Appl. Phys.} \textbf{2012}, \emph{51}, 06FF14\relax
\mciteBstWouldAddEndPuncttrue
\mciteSetBstMidEndSepPunct{\mcitedefaultmidpunct}
{\mcitedefaultendpunct}{\mcitedefaultseppunct}\relax
\EndOfBibitem
\bibitem[Lan \latin{et~al.}(2024)Lan, Yan, Yu, and Shen]{10.1021/acsami.4c03418}
Lan,~M.; Yan,~G.; Yu,~W.; Shen,~S. {Oxygen Impurity-Tuned Structure and Adhesion Properties of the Cu/SiO$_2$ Interface}. \emph{ACS Appl. Mater. Interfaces} \textbf{2024}, \emph{16}, 22724--22735\relax
\mciteBstWouldAddEndPuncttrue
\mciteSetBstMidEndSepPunct{\mcitedefaultmidpunct}
{\mcitedefaultendpunct}{\mcitedefaultseppunct}\relax
\EndOfBibitem
\bibitem[Deringer \latin{et~al.}(2019)Deringer, Caro, and Csányi]{10.1002/adma.201902765}
Deringer,~V.~L.; Caro,~M.~A.; Csányi,~G. {Machine Learning Interatomic Potentials as Emerging Tools for Materials Science}. \emph{Adv. Mater.} \textbf{2019}, \emph{31}, 1902765\relax
\mciteBstWouldAddEndPuncttrue
\mciteSetBstMidEndSepPunct{\mcitedefaultmidpunct}
{\mcitedefaultendpunct}{\mcitedefaultseppunct}\relax
\EndOfBibitem
\bibitem[Hong \latin{et~al.}(2023)Hong, Kim, Kim, Jung, Ju, Choi, and Han]{10.1080/27660400.2023.2269948}
Hong,~C.; Kim,~J.; Kim,~J.; Jung,~J.; Ju,~S.; Choi,~J.~M.; Han,~S. {Applications and training sets of machine learning potentials}. \emph{Sci. Technol. Adv. Mater. Methods} \textbf{2023}, \emph{3}, 2269948\relax
\mciteBstWouldAddEndPuncttrue
\mciteSetBstMidEndSepPunct{\mcitedefaultmidpunct}
{\mcitedefaultendpunct}{\mcitedefaultseppunct}\relax
\EndOfBibitem
\bibitem[Xu \latin{et~al.}(2025)Xu, Wu, Guo, Zhang, Zhong, Li, and Ren]{10.1063/5.0244175}
Xu,~S.; Wu,~J.; Guo,~Y.; Zhang,~Q.; Zhong,~X.; Li,~J.; Ren,~W. {Applications of machine learning in surfaces and interfaces}. \emph{Chem. Phys. Rev.} \textbf{2025}, \emph{6}, 011309\relax
\mciteBstWouldAddEndPuncttrue
\mciteSetBstMidEndSepPunct{\mcitedefaultmidpunct}
{\mcitedefaultendpunct}{\mcitedefaultseppunct}\relax
\EndOfBibitem
\bibitem[Cho \latin{et~al.}(2023)Cho, Son, Cho, Jang, Kim, and Min]{10.1038/s41598-023-44265-6}
Cho,~E.; Son,~W.-J.; Cho,~E.; Jang,~I.; Kim,~D.~S.; Min,~K. {Atomistic insights into adhesion characteristics of tungsten on titanium nitride using steered molecular dynamics with machine learning interatomic potential}. \emph{Sci. Rep.} \textbf{2023}, \emph{13}, 17145\relax
\mciteBstWouldAddEndPuncttrue
\mciteSetBstMidEndSepPunct{\mcitedefaultmidpunct}
{\mcitedefaultendpunct}{\mcitedefaultseppunct}\relax
\EndOfBibitem
\bibitem[Shapeev(2016)]{doi:10.1137/15M1054183}
Shapeev,~A.~V. Moment Tensor Potentials: A Class of Systematically Improvable Interatomic Potentials. \emph{Multiscale Model. Simul.} \textbf{2016}, \emph{14}, 1153--1173\relax
\mciteBstWouldAddEndPuncttrue
\mciteSetBstMidEndSepPunct{\mcitedefaultmidpunct}
{\mcitedefaultendpunct}{\mcitedefaultseppunct}\relax
\EndOfBibitem
\bibitem[Cho \latin{et~al.}(2025)Cho, Son, Lee, Do, Min, and Kim]{10.1039/d4tc04870a}
Cho,~E.; Son,~W.-J.; Lee,~S.; Do,~H.-S.; Min,~K.; Kim,~D.~S. {Unraveling the adhesion characteristics of ruthenium as an advanced metal interconnect material using machine learning potential}. \emph{J. Mater. Chem. C} \textbf{2025}, \emph{13}, 7772--7784\relax
\mciteBstWouldAddEndPuncttrue
\mciteSetBstMidEndSepPunct{\mcitedefaultmidpunct}
{\mcitedefaultendpunct}{\mcitedefaultseppunct}\relax
\EndOfBibitem
\bibitem[Qi \latin{et~al.}(2024)Qi, Sun, Sun, Wang, Zhang, Liang, Li, Zou, Li, Wu, Shen, and Liu]{10.1021/acsami.4c06055}
Qi,~Z.; Sun,~X.; Sun,~Z.; Wang,~Q.; Zhang,~D.; Liang,~K.; Li,~R.; Zou,~D.; Li,~L.; Wu,~G.; Shen,~W.; Liu,~S. {Interfacial Optimization for AlN/Diamond Heterostructures via Machine Learning Potential Molecular Dynamics Investigation of the Mechanical Properties}. \emph{ACS Appl. Mater. Interfaces} \textbf{2024}, \emph{16}, 27998--28007\relax
\mciteBstWouldAddEndPuncttrue
\mciteSetBstMidEndSepPunct{\mcitedefaultmidpunct}
{\mcitedefaultendpunct}{\mcitedefaultseppunct}\relax
\EndOfBibitem
\bibitem[Fan \latin{et~al.}(2021)Fan, Zeng, Zhang, Wang, Song, Dong, Chen, and Ala-Nissila]{PhysRevB.104.104309}
Fan,~Z.; Zeng,~Z.; Zhang,~C.; Wang,~Y.; Song,~K.; Dong,~H.; Chen,~Y.; Ala-Nissila,~T. Neuroevolution machine learning potentials: Combining high accuracy and low cost in atomistic simulations and application to heat transport. \emph{Phys. Rev. B} \textbf{2021}, \emph{104}, 104309\relax
\mciteBstWouldAddEndPuncttrue
\mciteSetBstMidEndSepPunct{\mcitedefaultmidpunct}
{\mcitedefaultendpunct}{\mcitedefaultseppunct}\relax
\EndOfBibitem
\bibitem[Jeong \latin{et~al.}(2025)Jeong, Sun, Cho, and Min]{10.1016/j.apsusc.2025.162558}
Jeong,~J.; Sun,~J.; Cho,~E.; Min,~K. {Unveiling strain-dependent adhesion behavior at TaN-Cu interface using machine learning interatomic potential}. \emph{Appl. Surf. Sci.} \textbf{2025}, \emph{689}, 162558\relax
\mciteBstWouldAddEndPuncttrue
\mciteSetBstMidEndSepPunct{\mcitedefaultmidpunct}
{\mcitedefaultendpunct}{\mcitedefaultseppunct}\relax
\EndOfBibitem
\bibitem[Deng \latin{et~al.}(2023)Deng, Zhong, Jun, Riebesell, Han, Bartel, and Ceder]{10.1038/s42256-023-00716-3}
Deng,~B.; Zhong,~P.; Jun,~K.; Riebesell,~J.; Han,~K.; Bartel,~C.~J.; Ceder,~G. {CHGNet as a pretrained universal neural network potential for charge-informed atomistic modelling}. \emph{Nat. Mach. Intell.} \textbf{2023}, \emph{5}, 1031--1041\relax
\mciteBstWouldAddEndPuncttrue
\mciteSetBstMidEndSepPunct{\mcitedefaultmidpunct}
{\mcitedefaultendpunct}{\mcitedefaultseppunct}\relax
\EndOfBibitem
\bibitem[Dulkin \latin{et~al.}(2011)Dulkin, Ko, Wu, Karim, Leeser, Park, Meng, and Ruzic]{10.1116/1.3602079}
Dulkin,~A.; Ko,~E.; Wu,~L.; Karim,~I.; Leeser,~K.; Park,~K.~J.; Meng,~L.; Ruzic,~D.~N. {Improving the quality of barrier/seed interface by optimizing physical vapor deposition of Cu Film in hollow cathode magnetron}. \emph{J. Vac. Sci. Technol. A} \textbf{2011}, \emph{29}, 041514\relax
\mciteBstWouldAddEndPuncttrue
\mciteSetBstMidEndSepPunct{\mcitedefaultmidpunct}
{\mcitedefaultendpunct}{\mcitedefaultseppunct}\relax
\EndOfBibitem
\bibitem[Jeong \latin{et~al.}(2025)Jeong, Kim, Park, Kim, Jang, Taura, Kokaze, Lee, and Yeom]{JEONG2025112307}
Jeong,~B.~H.; Kim,~D.~W.; Park,~D.~H.; Kim,~S.; Jang,~Y.~S.; Taura,~Y.; Kokaze,~Y.; Lee,~S.~H.; Yeom,~G.~Y. High-performance copper-seed-layer deposition using 60-MHz high-frequency–direct current superimposed magnetron sputtering. \emph{Microelectron. Eng.} \textbf{2025}, \emph{297}, 112307\relax
\mciteBstWouldAddEndPuncttrue
\mciteSetBstMidEndSepPunct{\mcitedefaultmidpunct}
{\mcitedefaultendpunct}{\mcitedefaultseppunct}\relax
\EndOfBibitem
\bibitem[Kresse and Furthmüller(1996)Kresse, and Furthmüller]{10.1016/0927-0256(96)00008-0}
Kresse,~G.; Furthmüller,~J. {Efficiency of ab-initio total energy calculations for metals and semiconductors using a plane-wave basis set}. \emph{Comput. Mater. Sci.} \textbf{1996}, \emph{6}, 15--50\relax
\mciteBstWouldAddEndPuncttrue
\mciteSetBstMidEndSepPunct{\mcitedefaultmidpunct}
{\mcitedefaultendpunct}{\mcitedefaultseppunct}\relax
\EndOfBibitem
\bibitem[Kresse and Furthmüller(1996)Kresse, and Furthmüller]{10.1103/physrevb.54.11169}
Kresse,~G.; Furthmüller,~J. {Efficient iterative schemes for ab initio total-energy calculations using a plane-wave basis set}. \emph{Phys. Rev. B} \textbf{1996}, \emph{54}, 11169--11186\relax
\mciteBstWouldAddEndPuncttrue
\mciteSetBstMidEndSepPunct{\mcitedefaultmidpunct}
{\mcitedefaultendpunct}{\mcitedefaultseppunct}\relax
\EndOfBibitem
\bibitem[Kresse and Joubert(1998)Kresse, and Joubert]{10.1103/physrevb.59.1758}
Kresse,~G.; Joubert,~D. {From ultrasoft pseudopotentials to the projector augmented-wave method}. \emph{Phys. Rev. B} \textbf{1998}, \emph{59}, 1758--1775\relax
\mciteBstWouldAddEndPuncttrue
\mciteSetBstMidEndSepPunct{\mcitedefaultmidpunct}
{\mcitedefaultendpunct}{\mcitedefaultseppunct}\relax
\EndOfBibitem
\bibitem[Kresse(1995)]{10.1016/0022-3093(95)00355-x}
Kresse,~G. {Ab initio molecular dynamics for liquid metals}. \emph{J. Non-Cryst. Solids} \textbf{1995}, \emph{192-193}, 222--229\relax
\mciteBstWouldAddEndPuncttrue
\mciteSetBstMidEndSepPunct{\mcitedefaultmidpunct}
{\mcitedefaultendpunct}{\mcitedefaultseppunct}\relax
\EndOfBibitem
\bibitem[Perdew \latin{et~al.}(1996)Perdew, Burke, and Ernzerhof]{10.1103/physrevlett.77.3865}
Perdew,~J.~P.; Burke,~K.; Ernzerhof,~M. {Generalized Gradient Approximation Made Simple}. \emph{Phys. Rev. Lett.} \textbf{1996}, \emph{77}, 3865--3868\relax
\mciteBstWouldAddEndPuncttrue
\mciteSetBstMidEndSepPunct{\mcitedefaultmidpunct}
{\mcitedefaultendpunct}{\mcitedefaultseppunct}\relax
\EndOfBibitem
\bibitem[Hellenbrandt(2004)]{10.1080/08893110410001664882}
Hellenbrandt,~M. {The Inorganic Crystal Structure Database (ICSD)—Present and Future}. \emph{Crystallogr. Rev.} \textbf{2004}, \emph{10}, 17--22\relax
\mciteBstWouldAddEndPuncttrue
\mciteSetBstMidEndSepPunct{\mcitedefaultmidpunct}
{\mcitedefaultendpunct}{\mcitedefaultseppunct}\relax
\EndOfBibitem
\bibitem[Wu \latin{et~al.}(2003)Wu, Ou, Chou, and Wu]{10.1149/1.1531974}
Wu,~W.-F.; Ou,~K.-L.; Chou,~C.-P.; Wu,~C.-C. {Effects of Nitrogen Plasma Treatment on Tantalum Diffusion Barriers in Copper Metallization}. \emph{J. Electrochem. Soc.} \textbf{2003}, \emph{150}, G83--G89\relax
\mciteBstWouldAddEndPuncttrue
\mciteSetBstMidEndSepPunct{\mcitedefaultmidpunct}
{\mcitedefaultendpunct}{\mcitedefaultseppunct}\relax
\EndOfBibitem
\bibitem[Thompson \latin{et~al.}(2022)Thompson, Aktulga, Berger, Bolintineanu, Brown, Crozier, {in 't Veld}, Kohlmeyer, Moore, Nguyen, Shan, Stevens, Tranchida, Trott, and Plimpton]{THOMPSON2022108171}
Thompson,~A.~P.; Aktulga,~H.~M.; Berger,~R.; Bolintineanu,~D.~S.; Brown,~W.~M.; Crozier,~P.~S.; {in 't Veld},~P.~J.; Kohlmeyer,~A.; Moore,~S.~G.; Nguyen,~T.~D.; Shan,~R.; Stevens,~M.~J.; Tranchida,~J.; Trott,~C.; Plimpton,~S.~J. LAMMPS - a flexible simulation tool for particle-based materials modeling at the atomic, meso, and continuum scales. \emph{Comput. Phys. Commun.} \textbf{2022}, \emph{271}, 108171\relax
\mciteBstWouldAddEndPuncttrue
\mciteSetBstMidEndSepPunct{\mcitedefaultmidpunct}
{\mcitedefaultendpunct}{\mcitedefaultseppunct}\relax
\EndOfBibitem
\bibitem[{Nos{\'e}}(1984)]{1984MolPh..52..255N}
{Nos{\'e}},~S. {A molecular dynamics method for simulations in the canonical ensemble}. \emph{Mol. Phys.} \textbf{1984}, \emph{52}, 255--268\relax
\mciteBstWouldAddEndPuncttrue
\mciteSetBstMidEndSepPunct{\mcitedefaultmidpunct}
{\mcitedefaultendpunct}{\mcitedefaultseppunct}\relax
\EndOfBibitem
\bibitem[Schneider and Stoll(1978)Schneider, and Stoll]{PhysRevB.17.1302}
Schneider,~T.; Stoll,~E. Molecular-dynamics study of a three-dimensional one-component model for distortive phase transitions. \emph{Phys. Rev. B} \textbf{1978}, \emph{17}, 1302--1322\relax
\mciteBstWouldAddEndPuncttrue
\mciteSetBstMidEndSepPunct{\mcitedefaultmidpunct}
{\mcitedefaultendpunct}{\mcitedefaultseppunct}\relax
\EndOfBibitem
\bibitem[Stukowski(2009)]{Stukowski_2010}
Stukowski,~A. Visualization and analysis of atomistic simulation data with OVITO–the Open Visualization Tool. \emph{Model. Simul. Mater. Sci. Eng.} \textbf{2009}, \emph{18}, 015012\relax
\mciteBstWouldAddEndPuncttrue
\mciteSetBstMidEndSepPunct{\mcitedefaultmidpunct}
{\mcitedefaultendpunct}{\mcitedefaultseppunct}\relax
\EndOfBibitem
\bibitem[Park \latin{et~al.}(2024)Park, Kim, Hwang, and Han]{10.1021/acs.jctc.4c00190}
Park,~Y.; Kim,~J.; Hwang,~S.; Han,~S. Scalable Parallel Algorithm for Graph Neural Network Interatomic Potentials in Molecular Dynamics Simulations. \emph{J. Chem. Theory Comput.} \textbf{2024}, \emph{20}, 4857--4868\relax
\mciteBstWouldAddEndPuncttrue
\mciteSetBstMidEndSepPunct{\mcitedefaultmidpunct}
{\mcitedefaultendpunct}{\mcitedefaultseppunct}\relax
\EndOfBibitem
\bibitem[Batzner \latin{et~al.}(2022)Batzner, Musaelian, Sun, Geiger, Mailoa, Kornbluth, Molinari, Smidt, and Kozinsky]{10.1038/s41467-022-29939-5}
Batzner,~S.; Musaelian,~A.; Sun,~L.; Geiger,~M.; Mailoa,~J.~P.; Kornbluth,~M.; Molinari,~N.; Smidt,~T.~E.; Kozinsky,~B. {E(3)-equivariant graph neural networks for data-efficient and accurate interatomic potentials}. \emph{Nat. Commun.} \textbf{2022}, \emph{13}, 2453\relax
\mciteBstWouldAddEndPuncttrue
\mciteSetBstMidEndSepPunct{\mcitedefaultmidpunct}
{\mcitedefaultendpunct}{\mcitedefaultseppunct}\relax
\EndOfBibitem
\bibitem[Lee \latin{et~al.}(2025)Lee, Kim, Park, Jeong, Han, Park, and Lee]{leeflashtp}
Lee,~S.~Y.; Kim,~H.; Park,~Y.; Jeong,~D.; Han,~S.; Park,~Y.; Lee,~J.~W. FlashTP: Fused, Sparsity-Aware Tensor Product for Machine Learning Interatomic Potentials. Forty-second International Conference on Machine Learning. 2025\relax
\mciteBstWouldAddEndPuncttrue
\mciteSetBstMidEndSepPunct{\mcitedefaultmidpunct}
{\mcitedefaultendpunct}{\mcitedefaultseppunct}\relax
\EndOfBibitem
\bibitem[Jarzynski(1997)]{PhysRevLett.78.2690}
Jarzynski,~C. Nonequilibrium Equality for Free Energy Differences. \emph{Phys. Rev. Lett.} \textbf{1997}, \emph{78}, 2690--2693\relax
\mciteBstWouldAddEndPuncttrue
\mciteSetBstMidEndSepPunct{\mcitedefaultmidpunct}
{\mcitedefaultendpunct}{\mcitedefaultseppunct}\relax
\EndOfBibitem
\bibitem[Zhang \latin{et~al.}(2012)Zhang, Liu, Shu, and Fan]{10.1016/j.apsusc.2012.08.082}
Zhang,~J.; Liu,~C.; Shu,~Y.; Fan,~J. {Growth and properties of Cu thin film deposited on Si(001) substrate: A molecular dynamics simulation study}. \emph{Appl. Surf. Sci.} \textbf{2012}, \emph{261}, 690--696\relax
\mciteBstWouldAddEndPuncttrue
\mciteSetBstMidEndSepPunct{\mcitedefaultmidpunct}
{\mcitedefaultendpunct}{\mcitedefaultseppunct}\relax
\EndOfBibitem
\bibitem[Park and Schulten(2004)Park, and Schulten]{park2004calculating-fbb}
Park,~S.; Schulten,~K. Calculating potentials of mean force from steered molecular dynamics simulations. \emph{J. Chem. Phys.} \textbf{2004}, \emph{120}, 5946--5961\relax
\mciteBstWouldAddEndPuncttrue
\mciteSetBstMidEndSepPunct{\mcitedefaultmidpunct}
{\mcitedefaultendpunct}{\mcitedefaultseppunct}\relax
\EndOfBibitem
\bibitem[Radisic \latin{et~al.}(2003)Radisic, Cao, Taephaisitphongse, West, and Searson]{Radisic_2003}
Radisic,~A.; Cao,~Y.; Taephaisitphongse,~P.; West,~A.~C.; Searson,~P.~C. Direct Copper Electrodeposition on TaN Barrier Layers. \emph{J. Electrochem. Soc.} \textbf{2003}, \emph{150}, C362\relax
\mciteBstWouldAddEndPuncttrue
\mciteSetBstMidEndSepPunct{\mcitedefaultmidpunct}
{\mcitedefaultendpunct}{\mcitedefaultseppunct}\relax
\EndOfBibitem
\bibitem[Yang and Chen(2003)Yang, and Chen]{Yang_2003}
Yang,~C.-Y.; Chen,~J.~S. Investigation of Copper Agglomeration at Elevated Temperatures. \emph{J. Electrochem. Soc.} \textbf{2003}, \emph{150}, G826\relax
\mciteBstWouldAddEndPuncttrue
\mciteSetBstMidEndSepPunct{\mcitedefaultmidpunct}
{\mcitedefaultendpunct}{\mcitedefaultseppunct}\relax
\EndOfBibitem
\bibitem[Li \latin{et~al.}(2020)Li, Tian, Teng, and Cao]{10.3390/ma13215049}
Li,~Z.; Tian,~Y.; Teng,~C.; Cao,~H. {Recent Advances in Barrier Layer of Cu Interconnects}. \emph{Materials} \textbf{2020}, \emph{13}, 5049\relax
\mciteBstWouldAddEndPuncttrue
\mciteSetBstMidEndSepPunct{\mcitedefaultmidpunct}
{\mcitedefaultendpunct}{\mcitedefaultseppunct}\relax
\EndOfBibitem
\bibitem[Chong \latin{et~al.}(2005)Chong, Ee, Chen, and Law]{10.1016/j.surfcoat.2004.10.086}
Chong,~S.; Ee,~Y.; Chen,~Z.; Law,~S. {Electroless copper seed layer deposition on tantalum nitride barrier film}. \emph{Surf. Coat. Technol.} \textbf{2005}, \emph{198}, 287--290\relax
\mciteBstWouldAddEndPuncttrue
\mciteSetBstMidEndSepPunct{\mcitedefaultmidpunct}
{\mcitedefaultendpunct}{\mcitedefaultseppunct}\relax
\EndOfBibitem
\bibitem[Sangiovanni(2018)]{SANGIOVANNI2018180}
Sangiovanni,~D. Copper adatom, admolecule transport, and island nucleation on TiN(001) via ab initio molecular dynamics. \emph{Appl. Surf. Sci.} \textbf{2018}, \emph{450}, 180--189\relax
\mciteBstWouldAddEndPuncttrue
\mciteSetBstMidEndSepPunct{\mcitedefaultmidpunct}
{\mcitedefaultendpunct}{\mcitedefaultseppunct}\relax
\EndOfBibitem
\bibitem[Nies \latin{et~al.}(2021)Nies, Natarajan, and Nolan]{10.1039/d1sc04708f}
Nies,~C.-L.; Natarajan,~S.~K.; Nolan,~M. {Control of the Cu morphology on Ru-passivated and Ru-doped TaN surfaces – promoting growth of 2D conducting copper for CMOS interconnects}. \emph{Chem. Sci.} \textbf{2021}, \emph{13}, 713--725\relax
\mciteBstWouldAddEndPuncttrue
\mciteSetBstMidEndSepPunct{\mcitedefaultmidpunct}
{\mcitedefaultendpunct}{\mcitedefaultseppunct}\relax
\EndOfBibitem
\bibitem[Han \latin{et~al.}(2010)Han, Wu, Zhou, Chen, Gordon, Lei, Roberts, and Cheng]{10.1002/ange.200905360}
Han,~B.; Wu,~J.; Zhou,~C.; Chen,~B.; Gordon,~R.; Lei,~X.; Roberts,~D.~A.; Cheng,~H. {First‐Principles Simulations of Conditions of Enhanced Adhesion Between Copper and TaN(111) Surfaces Using a Variety of Metallic Glue Materials}. \emph{Angew. Chem.} \textbf{2010}, \emph{122}, 152--156\relax
\mciteBstWouldAddEndPuncttrue
\mciteSetBstMidEndSepPunct{\mcitedefaultmidpunct}
{\mcitedefaultendpunct}{\mcitedefaultseppunct}\relax
\EndOfBibitem
\bibitem[Aldana \latin{et~al.}(2025)Aldana, Nies, and Nolan]{Aldana_2025}
Aldana,~S.; Nies,~C.-L.; Nolan,~M. Control of Cu morphology on TaN barrier and combined Ru-TaN barrier/liner substrates for nanoscale interconnects from atomistic kinetic Monte Carlo simulations. \emph{Nanoscale} \textbf{2025}, \emph{17}, 12450–12464\relax
\mciteBstWouldAddEndPuncttrue
\mciteSetBstMidEndSepPunct{\mcitedefaultmidpunct}
{\mcitedefaultendpunct}{\mcitedefaultseppunct}\relax
\EndOfBibitem
\bibitem[Stukowski(2014)]{10.1007/s11837-013-0827-5}
Stukowski,~A. {Computational Analysis Methods in Atomistic Modeling of Crystals}. \emph{JOM} \textbf{2014}, \emph{66}, 399--407\relax
\mciteBstWouldAddEndPuncttrue
\mciteSetBstMidEndSepPunct{\mcitedefaultmidpunct}
{\mcitedefaultendpunct}{\mcitedefaultseppunct}\relax
\EndOfBibitem
\bibitem[Stukowski \latin{et~al.}(2012)Stukowski, Bulatov, and Arsenlis]{Stukowski_2012}
Stukowski,~A.; Bulatov,~V.~V.; Arsenlis,~A. Automated identification and indexing of dislocations in crystal interfaces. \emph{Model. Simul. Mater. Sci. Eng.} \textbf{2012}, \emph{20}, 085007\relax
\mciteBstWouldAddEndPuncttrue
\mciteSetBstMidEndSepPunct{\mcitedefaultmidpunct}
{\mcitedefaultendpunct}{\mcitedefaultseppunct}\relax
\EndOfBibitem
\bibitem[Rice and and(1974)Rice, and and]{Rice01011974}
Rice,~J.~R.; and,~R.~T. Ductile versus brittle behaviour of crystals. \emph{Philos. Mag.} \textbf{1974}, \emph{29}, 73--97\relax
\mciteBstWouldAddEndPuncttrue
\mciteSetBstMidEndSepPunct{\mcitedefaultmidpunct}
{\mcitedefaultendpunct}{\mcitedefaultseppunct}\relax
\EndOfBibitem
\bibitem[Pan and Rupert(2015)Pan, and Rupert]{10.1016/j.actamat.2015.02.012}
Pan,~Z.; Rupert,~T.~J. {Amorphous intergranular films as toughening structural features}. \emph{Acta Mater.} \textbf{2015}, \emph{89}, 205--214\relax
\mciteBstWouldAddEndPuncttrue
\mciteSetBstMidEndSepPunct{\mcitedefaultmidpunct}
{\mcitedefaultendpunct}{\mcitedefaultseppunct}\relax
\EndOfBibitem
\end{mcitethebibliography}

\cleardoublepage

\end{document}